\documentclass[aps,prb,twocolumn]{revtex4-2}
\usepackage[utf8]{inputenc}
\usepackage{amsmath}
\usepackage{amsfonts}
\usepackage{amssymb}
\usepackage{bbold}
\usepackage{graphicx}
\usepackage{lmodern}
\usepackage{xcolor}
\usepackage{hyperref}
\usepackage[left=2cm,right=2cm,top=2cm,bottom=2cm]{geometry}
\graphicspath{{./Figures/}}

\newcommand{\ket}[1]{\left|#1\right\rangle}

\newcommand{\mat}[4]{\begin{pmatrix}#1 & #2\\#3 & #4\end{pmatrix}}

\newcommand{\msem}{m}

\newcommand{\tz}{\tau_z}
\newcommand{\taus}{\tau}
\newcommand{\sx}{\sigma_x}
\newcommand{\sy}{\sigma_y}
\newcommand{\sz}{\sigma_z}

\newcommand{\spz}{s_z}

\begin{document}
\title{Chiral chains with two valleys and disorder of finite correlation length}
\author{Jean-Baptiste Touchais}
\author{Pascal Simon}
\author{Andrej Mesaros}
\email{andrej.mesaros@universite-paris-saclay.fr}

\affiliation{Universit\'e Paris-Saclay, CNRS, Laboratoire de Physique des Solides, 91405, Orsay, France}

\date{\today}

\begin{abstract}
In one-dimensional disordered systems with a chiral symmetry it is well-known that electrons at energy $E=0$ avoid localization and simultaneously exhibit a diverging density of states (DOS). For $N$ coupled chains with zero-correlation-length disorder, the diverging DOS remains for odd $N$, but a vanishing DOS is found for even $N$.
%These results are confirmed for the low-energy continuum theory with $N$ isotropically coupled linearly dispersing electrons having continuum white noise.
We use a thin spinless graphene nanotube with disordered Semenoff mass and disordered Haldane coupling to construct $N=2$ chiral chain models which at low energy have two linear band crossings at different momenta $\pm K$ (two valleys) and disorder with an arbitrary correlation length $\xi$ in units of lattice constant $a$. We find that the finite momentum $\pm K$ forces the disorder in one valley to depend on the disorder in the other valley, thus departing from known analytical results which assume having $N$ independent disorders (whatever their spatial correlation lengths). Our main numerical results show that for this inter-dependent mass disorder the DOS is also suppressed in the limit of strongly coupled valleys (lattice-white noise limit, $\xi/a=0$) and exhibits a non-trivial crossover as the valleys decouple ($\xi/a\gtrsim5$) into the DOS shapes of the $N=1$ continuum model with finite correlation length $\xi$. We also show that changing the intra-unit-cell geometry of the disordered Haldane coupling can tune the amount of inter-valley scattering yet at lowest energies it produces the decoupled-valley behavior ($N=1$) all the way down to lattice white noise.
\end{abstract}

\maketitle

\section{Introduction}
\label{sec:intro}
%%%%%%%%%%%%%%%%%%%%%%%%%%%%%%%%%%%%%%%%%%%%%%%%%%%
\begin{figure}
	\centering
	\includegraphics[width=0.49\textwidth]{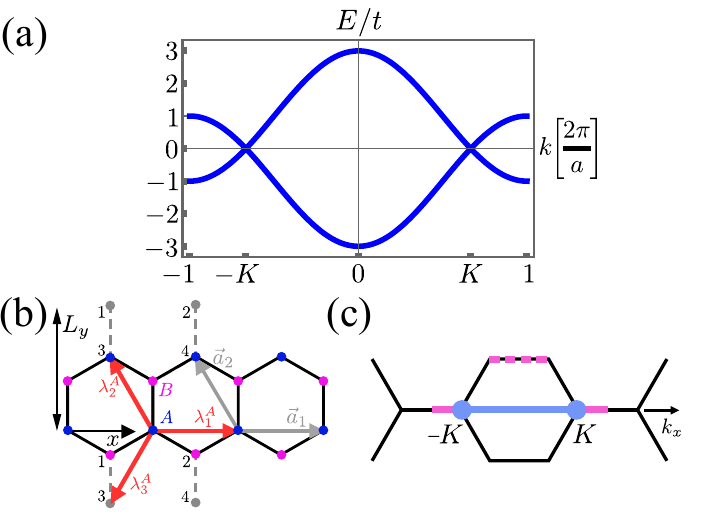}
	\caption{\label{fig:1} Quasi-one-dimensional models with chiral symmetry and on-site or bond disorder. (a) The clean lattice model without disorder has two bands, which in the continuum limit (energies near zero) form two one-dimensional Dirac electrons, at momenta $\pm K$, respectively. (b) The lattice model in real space. Numbered sites and dashed lines illustrate the periodic boundary conditions along $y$-axis. The on-site disorder term (mass) is given by the difference of on-site energies of $A$ and $B$ site in a unit-cell. The NNN-hopping disorder term is given by the three next-nearest-neighbor directed imaginary hoppings $\lambda_J^A$ emanating from site $A$ (red arrows). The $\lambda_j^B$ hoppings are obtained by translating from site $A$ to site $B$ above it and reversing direction. Note that $\lambda_2$ and $\lambda_3$ contribute to the same hopping term. (c) The momenta corresponding to system in (b) lie on the blue and dashed pink lines in the first Brillouin zone of infinite graphene. An equivalent description sets $k_y=0$ and extends the range of $k_x$ by the full pink lines.}
\end{figure}
%%%%%%%%%%%%%%%%%%%%%%%%%%%%%%%%%%%%%%%%%%%%%%%%%%%

In the decades following the discovery of localization of wavefunctions in one-dimensional disordered systems\cite{Anderson,Mott}, it became obvious that chiral symmetry, a non-spatial operation $\Xi$ for which the single-particle Hamiltonian $H$ obeys $\Xi H=-H\Xi$, enables delocalization of wavefunctions at the special energy $E=0$, with a simultaneously divergent Density of States (DOS) as $E\rightarrow 0$, first identified by Dyson\cite{Dyson}. Chiral symmetry occurs naturally in bipartite systems where the Hamiltonian maps one sublattice exclusively onto the other, and the Dyson peak was hence found in various physical contexts\cite{Shankar,Smith,DFisher,McCoy,Mathur,Furusaki} reducible to tight-binding chain models with a single site ($N=1$) in each unit-cell $i$ \cite{Gade,Theodorou,Eggarter} having disordered hoppings $t_{i,i+1}$ (odd/even $i$ forming the two sublattices). It suffices to consider \cite{Eggarter,Gade,Wegner} the limits of uncorrelated hoppings at each $i$, termed lattice-white noise disorder. The same diverging DOS was found in the low-energy continuum model of a disordered semiconductor\cite{Ovchinnikov}, due to the chiral symmetry of the single ($N=1$) two-component Dirac equation in one dimension with a disordered mass $m(x)$ obeying $\langle m(x)m(x')\rangle\propto m_0^2\delta(x-x')$ with $\delta(x)$ the Dirac delta function, i.e., $m(x)$ is continuum-white noise disorder. In condensed-matter contexts the system may often be quasi-one-dimensional ($N>1$ channels), and the disorder has a finite correlation length. The fate of the electronic DOS as the strength of disorder or the finite correlation length vary becomes less clear. 

A specific challenge relevant for condensed matter systems such as quantum wires, thin ribbons or nanotubes, occurs when the multiple channels come from valleys, such as in the minimal example of $N=2$ in Fig.~\ref{fig:1}a. In this case, a crucial new dimensionless parameter is $\xi/a$, the ratio of the correlation length of disorder and the lattice constant $a$. There are two main effects one must account for: (1) As we show below, the fact that valleys exist at finite momenta $\pm K$ implies that at low energies the disorder fields in the two valleys are not independent; and (2) As $\xi$ becomes comparable to $a$, the inter-valley scattering induces a crossover from two copies of $N=1$ to a single $N=2$ chiral system. The goal of this paper is to study the low-energy DOS for the entire range of $\xi/a$ and for weak and moderate strengths of disorder as quantified by the dimensionless parameter $m_0\xi/(\hbar v_F)$, where $v_F$ is the Fermi velocity in the valleys at zero energy, and hence construct the phase diagram in these two parameters.
%`N=2, a>0, X=xi/a, M=m*xi/(hv); E~m

Some limits in this phase diagram are understood and already show a variety of DOS behaviors in the non-diffusive regime. In particular, the electronic DOS due to a finite-correlation-length disorder was considered in detail for the single channel $N=1$ chiral chain\cite{Millis}. Correlated disorder was also considered for a Dirac equation with a spatially fluctuating gap, which covers many physical contexts \cite{BartoschThesis,CaoFGM} including the Peierls transition\cite{BartoschAN}. This model becomes chiral for real values of the gap\cite{BartoschAN,BartoschNUM,BartoschThesis}. All results were obtained in the continuum limit where $a\rightarrow 0, \xi/a\rightarrow\infty$. As the remaining dimensionless parameter $l\equiv\xi m_0/(\hbar v_F)$ is tuned by the length $\xi$, the evolution of DOS was found in detail\cite{BartoschNUM,BartoschExact}: The dip near the Dyson singularity (at $l=0$) deepens continuously as $l$ grows, converging to a pseudogap with DOS proportional to $E$ at small $E$. Nevertheless the singularity at $E=0$ persists at all finite $l$ under general boundary conditions\cite{BartoschComment,BartoschExact}. Hence in our two-valley system, when $\xi$ is sufficiently larger than $a$, the disorder is smooth enough and the inter-valley scattering vanishes, the two valleys form two independent copies of the fluctuating real gap model detailed in Ref.\onlinecite{BartoschThesis}, hence in this limit the problem is effectively $N=1$.

The most striking effect found in chiral multi-channel $N>1$ systems is the even-odd effect in $N$. \cite{Brouwer98N} Firstly, in the reference case $N=1$ mentioned above, the constant density of states of the clean system at low energies develops a small dip and then diverges at $E=0$ (the Dyson peak). \cite{Dyson,Ovchinnikov} Next, this phenomenology repeats for all odd $N$.\cite{BMF00PRL,BMF01physE,Grabsch} In stark contrast, for all even $N$ starting from $N=2$ the DOS vanishes\cite{BMF00PRL,BMF01physE} at $E=0$ and forms a hump at higher energies \cite{Grabsch}. The first important issue is that the disorder is assumed as white noise (either on lattice or in the continuum), i.e., $\xi/a,l\rightarrow 0$.
%leaving no dimensionless parameters (the average of m is taken as zero)
The second important issue is that in all these works the disorder is assumed to act isotropically in the channels, e.g., the intra-channel disorder fields $m_{aa}(x)$, $a=1\ldots N$ are independent white-noises. This assumption was motivated by a model of $N$ identical chains coupled in real space, and also by technical simplicity. In our case of $N=2$ valleys forming at finite momenta, the effective intra-valley disorder fields will turn out to be tied to each other, $m_{11}(x)=m_{22}(x)$, forming a sub-ensemble of the isotropic random fields. The mentioned works, especially the appealing closed-form solutions for DOS,\cite{Grabsch,GrabschThesis} are not adaptable in a straightforward way to this specific sub-ensemble. Hence, the known $N=2$ DOS\cite{BMF00PRL,Morimoto,Grabsch} is interesting for our two-valley system in the white noise limit $\xi/a<1$, but in principle does not apply.

These challenges motivated us to narrow the problem by assuming a vanishing average value of disorder, i.e., $\langle m_{ab}(x)\rangle=0$, which is a critical line in the $BDI$ Altland-Zirnbauer class at odd-$N$ with isotropic and white-noise disorder\cite{Morimoto}. For such disorder, the non-zero average completely removes the DOS singularity, which we aim to avoid in this work as the singularity is a striking feature that can be followed across the phase diagram. In terms of analyzing the DOS, it is important to note that the pseudogap-like form of DOS for $N=1$ at finite $l$ is both in principle and in practice distinguishable from the suppressed form of DOS in the $N=2$ model with $l=0$, hence they both provide characteristic behaviors to which our results may be compared.

To map out the two-valley phase diagram we choose a simple model that is derived from the familiar condensed matter system of graphene. We have a thin spinless graphene nanotube wrapped in the armchair direction (Fig.~\ref{fig:1}b), having the spectrum with two valleys in Fig.~\ref{fig:1}a, with chirality ensured by the minimal circumference of the tube. In order to emphasize the inter-valley scattering effects as $\xi/a$ varies we consider separately two types of disorders: in the Semenoff mass, and in the Haldane coupling, the latter inducing an a priori weaker inter-valley scattering because it is forbidden to back-scatter. With the Haldane coupling we can however tune the amount of inter-valley scattering by choosing the arrangement of next-nearest-neighbor hoppings inside the unit-cell.

This paper is organized as follows: In Section~\ref{sec:dis_mass_model} we consider the disordered Semenoff mass, first the lattice tight-binding model with disorder (Section~\ref{sec:dis_mass_model_latt}), then its low-energy continuum limit (Section~\ref{sec:dis_mass_model_cont}), and then in Section~\ref{sec:dis_mass_compare} we provide the mapping of parameters between the two, before we compare our continuum theory to the known models from literature and introduce their known DOS functions in Section~\ref{sec:known}. In Section~\ref{sec:dis_mass_num} we present the numerical results for DOS of our lattice model, and compare with the known DOS functions. In Section~\ref{sec:SOC} we consider a model with disordered NNN hoppings based on the Haldane coupling, first on the lattice and then in the continuum limit, with special attention brought to the connection of the intra-unit-cell structure of hoppings and the inter-valley scattering. In Section~\ref{sec:dis_soc_num} we present the numerically obtained DOS for the lattice model. We close with a discussion and conclusions.

\section{Disordered mass and two valleys}
\label{sec:dis_mass_model}
\subsection{Lattice model}
\label{sec:dis_mass_model_latt}
We begin by considering a spinless tight-binding model on the honeycomb lattice: It is a row of $N_x$ hexagonal plaquettes with periodic boundary conditions (PBC) in both directions (Fig.~\ref{fig:1}b). The model represents a very thin spinless graphene nanotube, i.e., it wraps a sample of graphene on a torus with PBC given by lattice translations $\vec{L}_x=N_x\vec{a}_1$ and $\vec{L}_y=2\vec{a}_2-\vec{a}_1$, where the $y$-axis is along the honeycomb armchair direction (see Fig.~\ref{fig:1}a). The tight-binding Hamiltonian is:
\begin{align}
  	\label{eq:Hmlattice}
	H_{m}^{TB} =& -t\displaystyle\sum_{<R\alpha,R'\alpha'>}c_{R\alpha}^\dagger c_{R'\alpha'}\\\notag
	&+\sum_{R} \msem(R)(c_{RA}^\dagger c_{RA}-c_{RB}^\dagger c_{RB})
\end{align}
where $c^\dagger_{R\alpha}$ creates an electron in unit-cell labeled by $x$-position $R=\left(x+\frac{1}{2}i\right)a$, with $a$ the lattice constant, on one of the two zig-zag chains $i=0,1$, and on sublattice $\alpha=A,B$. The $<r,r'>$ denotes nearest-neighbor sites with PBC applied (see Fig.~\ref{fig:1}b), while $t$ is the hopping energy. This quasi-one-dimensional Hamiltonian is simply obtained in momentum space by setting $k_y\equiv 0$ and applying the PBC in $x$-direction in graphene (see Fig.~\ref{fig:1}b,c).

The real-valued function $m(R)$ is the Semenoff mass in unit-cell $R$, and in a disordered system we assume it to have a finite correlation length $\xi$ and strength $m_0$, while its disorder-averaged value is fixed to zero throughout this paper:
\begin{align}
  \label{eq:massdis}
  &\langle m(R)m(R')\rangle=m_0^2\frac{a}{\sqrt{2\pi}\xi}\exp{\left[-\frac{(R-R')^2}{2\xi^2}\right]},\\
  &\langle m(R)\rangle=0.
\end{align}

To understand the localization physics of this quasi-1d system, we must first consider its symmetries. The lattice model Eq.~\eqref{eq:Hmlattice} has a chiral symmetry $\Sigma:\,c_{RA}\rightarrow i c_{RB},\,c_{RB}\rightarrow -i c_{RA}$ since thanks to the periodicity in $y$-direction a connection via $t$ between two unit-cells remains intact under the transformation. The model also has spinless time-reversal symmetry $T=K$, with $K$ the complex conjugation, and is in the $CI$ class. The model can be interpreted as having $N=2$ coupled chains, but there are two differences compared to the standard $N$-channel chiral systems such as studied in Ref.~\onlinecite{Brouwer98N}: (1) Our disorder is not lattice-white-noise but has a finite correlation length $\xi$, and (2) Our disorder field $m(R)$ is not as isotropic in the channels, which becomes obvious in the low-energy limit (Sec.~\ref{sec:dis_mass_model_cont}).

To interpret numerical results of DOS from $H_{m}^{TB}$ we are interested in the low-energy behavior of the model, particularly so that the disorder-free bandstructure can be linearized. The disorder strength $m_0$ must also remain sufficiently smaller than the kinetic energy given by $t$. We now proceed to take the low-energy continuum limit including the disorder.

\subsection{Low-energy one-dimensional Dirac model}
\label{sec:dis_mass_model_cont}
The two Dirac points of disorder-free graphene are projected to different momenta $k_x=\pm K$ of our system (see Fig.~\ref{fig:1}a), thus creating two valleys, because the $x$-axis is the zigzag direction. In presence of disorder, we derive the low energy continuum theory in real space by inverse Fourier transforming the small momenta around the Dirac points of Eq.~\eqref{eq:Hmlattice}, keeping all Fourier components that connect these small momenta within or between valleys, obtaining:
\begin{align}
  \label{eq:Hmcont}
  \mathcal{H}_m^{cont}=&\hbar v_F\tz\sx(-i\partial_x)+\\\notag
 &\overline{m}(x)\sz+\\\notag
    &\textrm{Re}[\hat{m}(x)]\tau_x\sz-\textrm{Im}[\hat{m}(x)]\tau_y\sz
\end{align} 
where the Pauli matrices $\sigma_i$ act in sublattice ($A/B$)space, while $\tau_i$ act in valley space, with $\tau_z=\pm1$ labeling $\pm K$. The first line of Eq.~\eqref{eq:Hmcont} is the disorder-free kinetic energy of graphene with $k_y=0$. Introducing the Fourier transform $\tilde{f}(k)\equiv\sum_R\, e^{-ikR}f(R)$ of a lattice function $f(R)$, we defined its small-momentum-filtered function $\overline{f}(x)\equiv\int_{|q|\ll K}\textrm{d}q\, e^{iqx}\tilde{f}(q)$ which determines the intra-valley scattering. Note that for us $\overline{m}(x)$ is real-valued because $m(R)$ is such. We further defined $\hat{f}(x)\equiv\int_{|q|\ll K}\textrm{d}q\, e^{iqx}\tilde{f}(2K+q)$ as the (complex) slow-field envelope of the $2K$ Fourier component which determines the inter-valley scattering (IVS). Obviously, these IVS-causing Fourier components arise from $m(R)$ of the lattice model only when $\xi$ becomes comparable to $a$ (Eq.~\eqref{eq:massdis}). Hence, the second and third line of $\mathcal{H}_m^{cont}$ describe intra- and inter-valley scattering, respectively, due to a spatially varying mass term.

The symmetry properties of the low energy model in Eq.~\eqref{eq:Hmcont} are crucial for understanding its localization physics. There is always the chiral symmetry $\Sigma=\sy$. Next, consider the situation when the two valleys are decoupled, $\hat{m}(x)\equiv0$. Then $\tau_z$ is a unitary symmetry, and $\mathcal{H}_m^{cont}$ reduces to two identical independent valleys, each having the effective time-reversal symmetry $T=\sz K$, which obeys $T^2=1$ and commutes with $\Sigma$, so each valley being in the chiral class $BDI$. In presence of IVS, $\hat{m}(x)\neq 0$, there are no unitary symmetries, and the true spinless time-reversal symmetry of two valleys is $T=\tau_x K$, which still obeys $T^2=1$, but now anticommutes with $\Sigma$, giving the chiral class $CI$, as for the full lattice model. The one-dimensional Dirac equation is hence always in a chiral class, and should realize a crossover from $N=2$ $CI$ (with IVS) to two copies of $N=1$ $BDI$ (without IVS) as $\xi/a$ grows.

\subsection{Comparison of lattice and continuum}
\label{sec:dis_mass_compare}
%\subsubsection{Lattice vs. continuum}
It is worth to precise the relationship between the continuum model (Sec.~\ref{sec:dis_mass_model_cont}) and the lattice model (Sec.~\ref{sec:dis_mass_model_latt}). The low-energy continuum model requires taking the lattice scale to be the smallest in the problem. Consequently, the correlation-length parameter $\xi_{latt}/a\rightarrow\infty$ of the lattice model disappears in the continuum limit. Our expressions in Eq.~\eqref{eq:Hmcont} show how both the intra- and inter-valley-scattering disorder fields in the continuum $\overline{m}(x)$, $\hat{m}(x)$ inherit finite strengths and correlation lengths from the lattice function $m(R)$.

The linearization of the disorder-free bandstructure gives
\begin{equation}
  \label{eq:latt-cont}
\hbar v_F=\frac{\sqrt{3}}{2}t a  
\end{equation}
as in graphene, hence the low-energy limit is more precisely $a\rightarrow0,\,t\rightarrow\infty$ leaving the finite scale $\hbar v_F$, so that in the continuum the correlation length is measured in units of the only lengthscale $h v_F/m_0$. The natural dimensionless parameter in the continuum is hence $\frac{m_0\xi_{cont}}{\hbar v_F}$, which by Eq.~\eqref{eq:latt-cont} is proportional to $\frac{m_0}{t}\frac{\xi_{cont}}{a}$. The latter form is straightforwardly interpretable on the lattice by replacing $\xi_{cont}$ by $\xi_{latt}$. Hence, to compare results of lattice calculations which use $\xi_{latt}$ with continuum calculations which use $\xi_{cont}$, we match the two forms of this dimensionless parameter:
\begin{equation}
  \label{eq:param_map}
\frac{m_0\xi_{cont}}{\hbar v_F}\equiv f_l\frac{\xi_{latt}}{a}\frac{m_0}{t},
\end{equation}
up to a numerical prefactor $f_l$. The $f_l$ according to Eq.~\eqref{eq:latt-cont} is $2/\sqrt{3}$, but in principle it is expected to be a quantity of order 1 which possibly weakly depends on the value of $\xi_{latt}m_0$. In comparing our lattice DOS with known DOS of continuum models, we find that a constant value $f_l=2$ works very well. For clarity of presentation, we will therefore omit writing the $f_l$ factor when comparing numerical and known DOS functions.

Considering the energy scales, the $t$ is the natural one on the lattice, but it in the continuum only an energy scale of the disorder strength exists. As we will discuss in detail with concrete DOS functions in Section~\ref{sec:known}, the continuum disorder energy scale should be matched simply with the lattice quantity $m_0$ when the correlation length is finite, but it must be matched with the lattice quantity $m_0\frac{m_0}{t}$ when the correlation length vanishes. Hence one should be aware that the continuum energy scale is emergent, and hence to match a lattice DOS $\rho_{latt}(E/m_0)$ to a continuum DOS function $\rho_{cont}$, one may need to use $\rho_{cont}(E/m^{eff}_0)$, with an effective scale
\begin{equation}
  \label{eq:f-factor}
  m^{eff}_0=g_m m_0,
\end{equation}
corrected be a numerical factor $g_m$. We again expect $g_m$ to be a quantity of order 1 that may weakly depend on $m_0$. We find that $g_m=1.5$ is appropriate in the regime having smaller values $\frac{\xi_{latt}}{a}\frac{m_0}{t}<10$, while the prefactor becomes trivial, $g_m=1$, in the regime having larger values. For clarity of presentation, we will omit writing the $g_m$ factor in comparing various DOS in the following.

The special case of continuum-white noise $\frac{\xi_{cont}}{\hbar v_F/m_0}\rightarrow 0$ (in which there are no remaining dimensionless parameters\cite{Ovchinnikov}) should be related to the case of lattice white noise $\xi_{latt}/a\rightarrow 0$ as long as the dimensionless parameter $m_0/t$ remains finite (for our goal of finding low-energy properties, this ratio is always less than 1).

The key properties of the low-energy disorder inherited from the lattice disorder become obvious by rewriting the Hamiltonian Eq.~\eqref{eq:Hmcont}:
\begin{align}
  \label{eq:HmcontM}
  &\mathcal{H}_m^{cont}=\hbar v_F\tz\sx(-i\partial_x)+M(x)\sz\\\notag
&M(x)=\mat{\overline{m}(x)}{\hat{m}(x)}{\hat{m}(x)^*}{\overline{m}(x)},
\end{align} 
showing that the disorder matrix necessarily exhibits an inter-dependence between valleys
\begin{equation}
  \label{eq:dis_corr}
  M_{11}(x)=M_{22}(x).
\end{equation}
It is important to note that our valleys form a complex $4\times4$ representation, hence even though we have (spinless) time-reversal symmetry, the ensemble of disorder matrices $M(x)$ relevant for us is complex\cite{Brouwer98N,Grabsch}. Only in the limit of vanishing IVS, $\hat{m}(x)\equiv0$, the disorder in each independent valley becomes an ensemble of real functions.

To better understand the implications of our disordered mass matrix $M(x)$ in the continuum model Eq.~\eqref{eq:Hmcont}, we apply transfer-matrix theory (Appendix~\ref{app:A}) to a general case where the noise has zero correlation length in the continuum, but there is an arbitrary amount of IVS ($\hat{m}(x)\neq0$) and the inter-dependence of disorders in valleys is respected (Eq.~\eqref{eq:dis_corr}). We find an insulator, with the same behavior of conductance as found in models of $N=2$ with independent noises (i.e., where the condition Eq.~\eqref{eq:dis_corr} is absent)\cite{Brouwer98N,BMF00PRL,Morimoto,Grabsch}. However, we cannot conclude that our low-energy DOS is suppressed at $E=0$, because the Thouless formula that connects the DOS and the conductance is not valid for $N>1$ \cite{BMF00PRL}.

\subsection{Models with known DOS}
\label{sec:known}
\subsubsection{N=2, independent white noises ($\rho^D_{N=2}(E)$)}
The general case $N>1$ was previously studied \cite{Brouwer98N,BMF00PRL,Morimoto,Grabsch} essentially under the assumption of a disorder matrix of independent isotropic white noises, i.e.,
\begin{equation}
  \label{eq:dis_iso}
\langle M_{ij}(x)^* M_{ij}(x')\rangle=D\delta(x-x'),\,\forall i,j.
\end{equation}
The low-energy Dirac equation with $N>1$ was connected to the lattice model of $N$ coupled chains in Ref.~\onlinecite{Brouwer98N}, finding for even $N$, such as $N=2$, an insulating state. The DOS at lowest energies was found to rise sharply from zero as $\rho(E)\sim |E\log(E)|$ (dubbed a ``pseudogap'')\cite{BMF00PRL,BMF01physE,Morimoto}.

A full analytical expression for the DOS in the continuum was derived for a disordered multi-channel Dirac equation in Ref.~\onlinecite{Grabsch}, using the following model (up to unitary rotations in $\sigma$ space):
\begin{equation}
\label{eq:grabsch_model}
\mathcal{H}=-i\sx\otimes\mathbf{1}_N\partial_x + \sz\otimes M(x)
\end{equation}
where $\mathbf{1}_N$ is the $N\times N$ identity matrix. Again, the disorder $M(x)$ is a random $N\times N$ hermitian matrix whose elements are independent Gaussian white noises, i.e., in the particular case of $N=2$ the diagonal terms $M_{11}(x)$ and $M_{22}(x)$ are independent. The DOS\cite{Grabsch} for the model in Eq.~\eqref{eq:grabsch_model} develops a hump at an energy which scales as
\begin{equation}
  \label{eq:grabsch_hump}
E^{N=2}_{hump}\sim D/(\hbar v_F),
\end{equation}
which is comparable to $m_0^2/t$ on the lattice (see Eqs.~\eqref{eq:massdis}, \eqref{eq:latt-cont}, \eqref{eq:dis_iso}). For higher energies the DOS relaxes to the disorder-free constant value. We label this particular DOS function from Ref.\cite{Grabsch} ($N=2$, complex matrices, vanishing average value of disorder), having the energy scale $D/(\hbar v_F)$, as $\rho^D_{N=2}(E)$.

In our model, the disorders in two valleys are not independent (Eqs.~\eqref{eq:HmcontM}, \eqref{eq:dis_corr}), so in presence of IVS we find that our disordered model is not unitarily equivalent to the above cited models. We were unable to generalize the methods that were used in Refs.\cite{BMF00PRL,Grabsch} to calculate the DOS for our type of inter-dependent disorders in the valleys,\footnote{Note, our kinetic term is $k_x\sx\tz$ while in Eq.~\eqref{eq:grabsch_model} it is $k_x\sx\mathbf{1}_2$, and these are also not unitarily equivalent. However, the method can be adapted to deal with this change, while in contrast we couldn't find a way to adapt it to deal with the different noise type.} even in this limit of white noises (zero correlation length).

\subsubsection{N=1, white noise ($\rho^D_{N=1}(E)$)}
In the seminal work\cite{Ovchinnikov} on the  $N=1$ problem of a Dirac particle with a single random mass field $m(x)$, the $m(x)$ is a continuum Gaussian white noise with
\begin{equation}
  \label{eq:ovch_dis}
  \langle m(x)m(x')\rangle=D\delta (x-x').
\end{equation}
Taking a vanishing average $\langle m(x)\rangle\equiv0$ relevant for us, we recover the exact analytical form for the integrated DOS:
\begin{equation}
\label{eq:ovchinnikov}
\mathcal{N}(E) = \frac{2D}{\hbar v_F\pi^2(J^2_{0}(\hbar v_F E/D)+Y^2_{0}(\hbar v_F E/D))} \,,
\end{equation}
where $J_{0}$ and $Y_{0}$ are the zero-th Bessel functions of the first and second kind, respectively. There are no dimensionless parameters, and we label the resulting DOS function $\frac{\textrm{d}\mathcal{N}(E)}{\textrm{d}E}\equiv\rho^D_{N=1}(E)$. 

The DOS at lowest energies has the Dyson singularity, $\rho(E)\sim1/|E\log(E)|^3$, then drops to form a dip of about $\delta\rho_{dip}\approx3\%$ (compared to the disorder-free constant value of DOS) positioned at energy $E_{dip}\sim D/(\hbar v_F)$ (this energy is comparable to $m_0^2/t$ on the lattice by Eq.~\eqref{eq:param_map}), before relaxing to the disorder-free constant value at higher energies.

In our model, we expect this DOS to be relevant in a corner of the phase diagram, where both the IVS vanishes (the valleys decouple into two copies of $N=1$), and simultaneously the disorder is white noise (vanishing continuum correlation length).

\subsubsection{N=1, noise with finite correlation length ($\rho^{\mu,l}_{N=1}(E)$)}
Refs.~\onlinecite{BartoschAN,BartoschNUM,BartoschThesis} expand on the result of Ovchinnikov (Ref.~\onlinecite{Ovchinnikov}) by considering an  $N=1$ Dirac equation with coloured noise, i.e., the mass $m(x)$ being correlated on a lengthscale $\xi_{dis}$:
\begin{equation}
  \label{eq:bart_dis}
  \langle m(x)m(x')\rangle=\mu^2\exp(-|x-x'|/\xi_{dis}),
\end{equation}
where we focus on the particular case of a real field $m(x)$ (in Ref.\onlinecite{BartoschNUM}, the case of vanishing phase fluctuations). The single dimensionless parameter $l\equiv \mu\xi_{dis}/(\hbar v_F)$ tunes the DOS from the continuum Gaussian white noise case of Ref.~\onlinecite{Ovchinnikov} (for which $l=0$) by monotonously deepening the dip $\delta\rho_{dip}(l)$ and moving its position $E_{dip}(l)\rightarrow0$. As $E_{dip}(l)$ moves to zero, the Dyson singularity is squeezed to $E=0$, while the DOS at lowest $E$ becomes linear $\rho(E)\sim E$, and then develops a hump whose position scales as $E^{N=1}_{hump}\sim \mu$ (comparable to $m_0$ on the lattice by Eq.~\eqref{eq:param_map}). We label this particular DOS function from Ref.\cite{BartoschNUM} (real mass, vanishing average value of disorder), having an energy scale $\mu$ and a fixed parameter $l$, as $\rho^{\mu,l}_{N=1}(E)$.

To be explicit, in the limit of continuum-white noise, $\xi_{dis}\rightarrow0$, one recovers the Ovchinnikov DOS in Eq.~\eqref{eq:ovchinnikov} by taking $l\rightarrow0$ simultaneously with $\mu\propto\frac{1}{\sqrt{\xi_{dis}}}$, so that $\rho^{\mu,l}_{N=1}(E)\rightarrow\rho^D_{N=1}(E)$ with $D\equiv\frac{\sqrt{2\pi}}{2}\mu^2\xi_{dis}$.

In our model, we expect the $\rho^{\mu,l}_{N=1}(E)$ DOS to be relevant in the limit $\xi_{latt}/a\gg1$ where the IVS vanishes (the valleys decouple into two copies of $N=1$), while the dimensionless correlation length parameter $l_{latt}\equiv\frac{\xi_{latt}}{a}\frac{m_0}{t}$ remains arbitrary.

\subsubsection{Summary of comparison to our model}
In summary, none of the known results apply strictly to our two-valley lattice model's entire phase diagram, which is controlled by the parameters $m_0\xi_{latt}/(t a)$ and $\xi_{latt}/a$ (Sec.~\ref{sec:dis_mass_model_latt}). However, based on the low-energy theories, we expect a meaningful comparison in a few limits, in particular:
\begin{enumerate}
\item At $\xi_{latt}/a\gg1$ we should recover the known $\rho^{\mu,l}_{N=1}(E)$ DOS by matching $l\equiv l_{latt}\equiv\xi_{latt}m_0/(t a)$, further $\mu\equiv m_0$, while we can in principle further test the scaling of $E^{N=1}_{hump}/m_0$, and of $E_{dip}(l_{latt})/m_0$ with disorder strength $m_0/t$ at fixed $l_{latt}$,
\item In the lattice-white-noise limit $\xi_{latt}<a$, the DOS depends only on $m_0/t$ (note, in this limit $l_{latt}$ reduces to this quantity), so we may attempt to compare to the known $\rho^D_{N=2}(E)$ DOS by matching $D/(\hbar v_F)\equiv m_0^2/t$, i.e., matching $D\equiv m_0^2a$ (Eq.~\eqref{eq:latt-cont}). We can further test the scaling of $E^{N=2}_{hump}/m_0$ with $m_0/t$. However, due to the inter-dependence of our disorders in the valleys, \textit{a priori} there is no guarantee of a match even in this limit.
\end{enumerate}

\section{Numerical DOS for disordered mass on lattice}
\label{sec:dis_mass_num}
%- fig.3: cut m=const=low; NEW->O->B->FGM
%- fig.4: cut m=const=large; NEW->weird->B->FGM
\subsection{Numerical set-up}
We numerically diagonalize the two-valley lattice model Eq.~\eqref{eq:Hmlattice} of lengths $N_x=10^4-10^5$ with disordered mass of strength $m_0$ and finite correlation length $\xi_{latt}$, Eq.~\eqref{eq:massdis}, and average the resulting DOS over $10^2-10^3$ disorder realizations. The accessible part of the phase diagram is limited by four conditions: (1) The disorder strength has to surpass the level spacing for disorder to have an effect, hence $m_0/t>1/N_x$; (2) the disorder has to be weak enough so that the results of low-energy theory are applicable, hence $m_0/t<1$; (3) The system should contain plenty of real-space domains that appear on lengthscale $\xi$ so that the finite-correlation-length physics is correctly sampled, hence $\xi_{latt}\ll N_x a$; (4) For values $\xi_{latt}/a\leq0.5$ the mass is numerically indistinguishable from lattice white noise $\xi_{latt}=0$ because of the quick Gaussian decay in Eq.~\eqref{eq:massdis}. The conditions (1) and (2) are the slanted straight lines marked on the phase diagrams, Figs.~\ref{fig:mass}, \ref{fig:soc}.
%In labeling the parameters of the numerical calculations, we use the mapping Eq.~\eqref{eq:param_map} to express them in terms of quantities in the low-energy theory.

\subsection{Results in the extreme limits}
\begin{figure}\centering\includegraphics[width=0.48\textwidth]{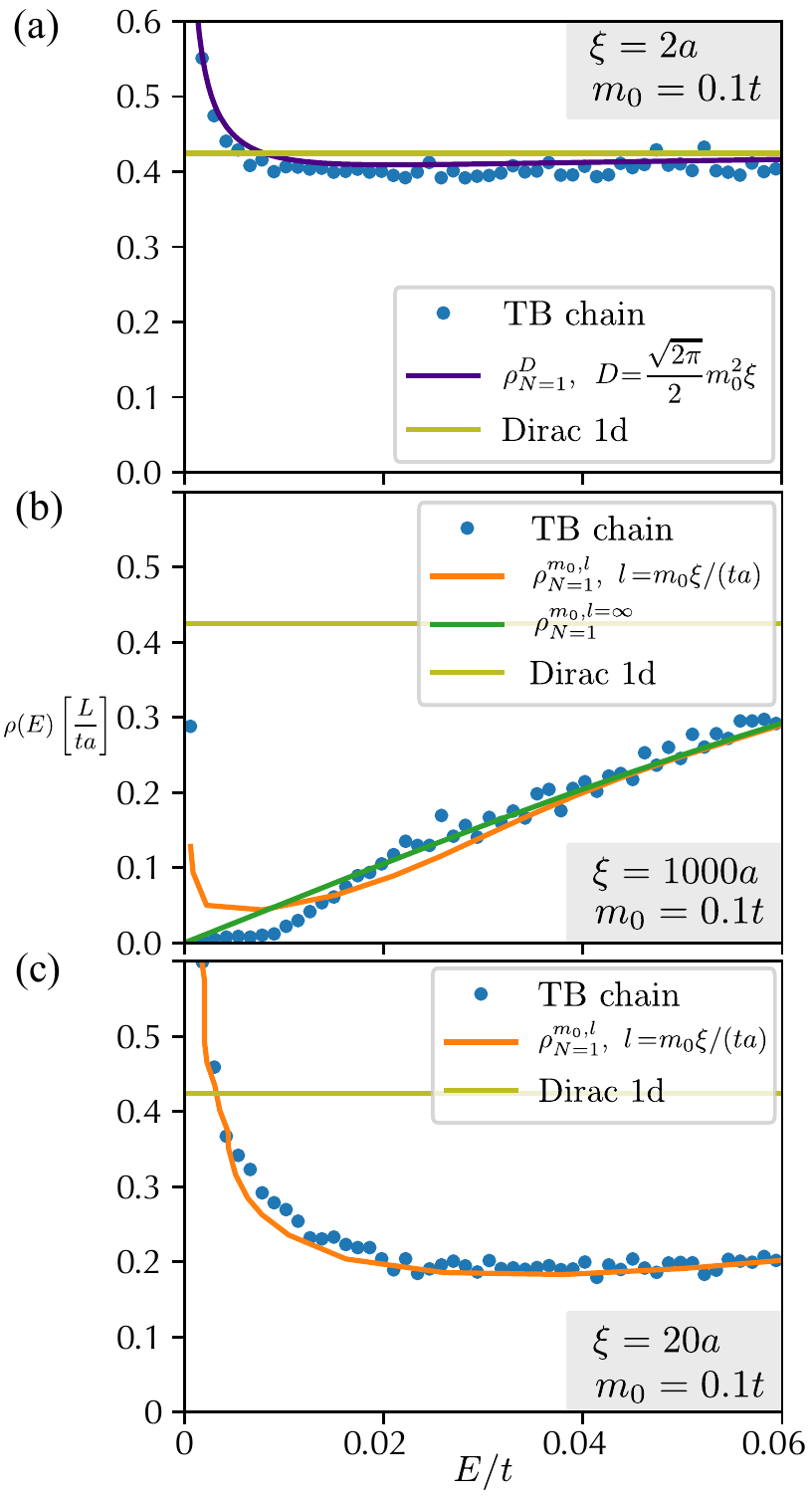}
\caption{\label{fig:masscut1} Density of states (DOS) of the two-valley model with disordered mass at extreme limits of vanishing inter-valley scattering (IVS). The green line in every figure is the DOS of disorder-free model. (a) For disorder-strength $m_0=0.1t$, at correlation length of only $\xi/a=2$ the inter-valley scattering is already suppressed and we have a good agreement (orange line) with the Ovchinnikov DOS $\rho^{\mu=m_0,l=0}_{N=1}(E)\equiv \rho^{D}_{N=1}(E)$ (see below Eq.~\eqref{eq:bart_dis}). Note, the lattice parameter $l_{latt}=0.2<1$ is indeed expected to give a quantitatively very similar DOS as $l=0$ (orange line). (b) For $m_0=0.1t$ and $\xi/a=1000$ we have $l_{latt}=100\gg 1$, and we find a reasonable agreement (orange line) with the $\rho^{\mu=m_0,l=100}_{N=1}(E/f)$. The data is near the limit of the linearly rising DOS $\rho^{\mu=m_0,l=\infty}_{N=1}(E)$ (dark green). (c) Intermediate situation between (a) and (b), with $m_0=0.1t$ and $\xi/a=20$ giving $l_{latt}=2$. We have a good agreement with $\rho^{\mu=m_0,l=1}_{N=1}(E)$ (orange line).}
\end{figure}

In Figure \ref{fig:masscut1} we illustrate that DOS curves in the extreme regime of negligible IVS ($\xi/a\gg1$).\footnote{The strict chirality of our model implies $\rho(-E)=\rho(E)$, confirmed numerically, so we only show $E\geq0$.} The DOS follows well the one-parameter scaling of $\rho^{\mu,l}_{N=1}(E)$ (see below Eq.~\eqref{eq:bart_dis}), at $l$ values up to 10 (Fig.\ref{fig:masscut1}a,c), while we need a quantitative correction in the energy scale such that $\rho(E)\rightarrow \rho(E/f)$  (see Eq.~\eqref{eq:f-factor}), with the example $f\approx0.67$ when $l=100$ (Fig.\ref{fig:masscut1}b). The divergence at $E=0$ is hard to track numerically, but we find very good quantitative agreement at finite energies including the dip feature.
%%%%%%%%%%%%%%%%%%%%%%%%%%%%%%%%%%%%%%%%%%%%%%%%%%%
%%%%%%%%%%%%%%%%%%%%%%%%%%%%%%%%%%%%%%%%%%%%%%%%%%%

On the other hand, when IVS is strongest (lattice white noise, $\xi/a=0$), in Fig.~\ref{fig:masscut2}a we show a surprising match with the DOS $\rho^D_{N=2}(E)$ (see below Eq.~\eqref{eq:grabsch_hump}). The logarithmic diverging slope at $E=0$ is hard to track in numerical data with finite resolution, but the hump feature is easily accessible. By tuning $m_0/t$ we confirm that the position of the hump follows $E_{hump}\sim m_0^2/t$ as is expected for $E^{N=2}_{hump}$ (Eq.~\eqref{eq:grabsch_hump}). Hence we uncover that the intrinsic inter-dependence of the disordered mass in the two-valleys (Eq.~\eqref{eq:dis_corr}) apparently does not affect the DOS at the lowest $m_0/t<0.2$. However, there are quantitative differences (similar to Fig.~\ref{fig:masscut2}b) for $\xi=0$ when $0.2<m_0/t<0.5$, which are disorder-strength values still low enough so that the regimes of low-energy Dirac particle and weak disorder should be valable. Further analytical investigations of the DOS of the two-valley system even in the lattice-white noise may hence be valuable.

\subsection{The two-parameter phase diagram}
\begin{figure}\centering\includegraphics[width=0.45\textwidth]{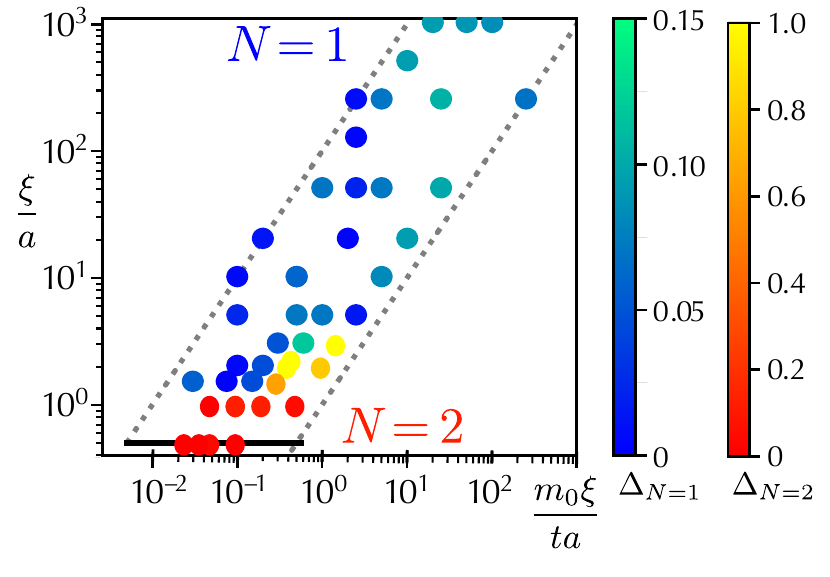}
\caption{\label{fig:mass} Phase diagram of the two-valley system with disordered mass. The green-to-blue scale measures the high-to-low agreement with $\rho^{\mu,l}_{N=1}(E)$ based on the depth of the dip in DOS, while the red-to-yellow measures the high-to-low agreement with $\rho^{D}_{N=2}(E)$ based on the height of the hump in DOS (see text for the definitions of $\Delta_N$). The dotted lines show the limits of our numerical calculations, while the thick line shows that all values $\xi/a\leq0.5$ produce essentially the same lattice disorder with zero correlation length.}
\end{figure}

In Fig.~\ref{fig:mass} we illustrate the entire phase diagram: at each point we qualitatively distinguish two types of numerically obtained lattice DOS: if its highest value is at the lowest available energy, we judge that it has a peak at $E=0$ and hence it is comparable to the known $\rho^{\mu,l}_{N=1}(E)$, while if the DOS has a peak (hump) at a finite energy it is comparable to $\rho^D_{N=2}(E)$. The former type is colored blue, the latter red, while we roughly quantify the agreement (high-to-low) by the shade (blue-to-green, and red-to-yellow, respectively). To quantify the agreement with $\rho^{\mu,l}_{N=1}(E)$ we compare the depth of the dip, i.e. with the known DOS value at the dip $\rho_{dip}\equiv \rho^{\mu=m_0,l=l_{latt}}_{N=1}(E_{dip}(l=l_{latt}))$ we define
\begin{equation}
  \label{eq:D1}
  \Delta_{N=1}\equiv|\rho_{min}-\rho_{dip}|/\rho_0,
\end{equation}
where $\rho_{min}$ is the minimum of the numerical DOS, and $\rho_0$ is the constant DOS value of the disorder-free low-energy Dirac theory. To quantify the agreement with $\rho^{D}_{N=2}(E)$, we measure the height of the hump, i.e., with the known DOS value at the hump $\rho_{hump}\equiv\rho^{D=m_0^2a}_{N=2}(E^{N=2}_{hump})$ we define
\begin{equation}
  \label{eq:D2}
\Delta_{N=2}\equiv|\rho_{max}-\rho_{hump}|/\rho_{hump},
\end{equation}
where $\rho_{max}$ is the maximum of the numerical DOS. We emphasize that these are rough quantifications of agreement, because the lattice DOS and the known DOS might not overlap well, and the energy value at which the dip (or hump) appears may differ. Certainly this is the case in the crossover region, but the goal of the rough quantification in the phase diagram Fig.~\ref{fig:mass} is exactly to make visible the crossover region.
%%%%%%%%%%%%%%%%%%%%%%%%%%%%%%%%%%%%%%%%%%%%%%%%%%%
%%%%%%%%%%%%%%%%%%%%%%%%%%%%%%%%%%%%%%%%%%%%%%%%%%%
%, which for $\xi/a\gg1$ and $l\rightarrow 0$ should be $\delta\rho_{dip}(l=0)=3\%$ (see after Eqs.~\eqref{eq:ovchinnikov},~\eqref{eq:bart_dis})

\subsection{Crossover between $N=2$ and $N=1$}
\begin{figure}\centering\includegraphics[width=0.45\textwidth]{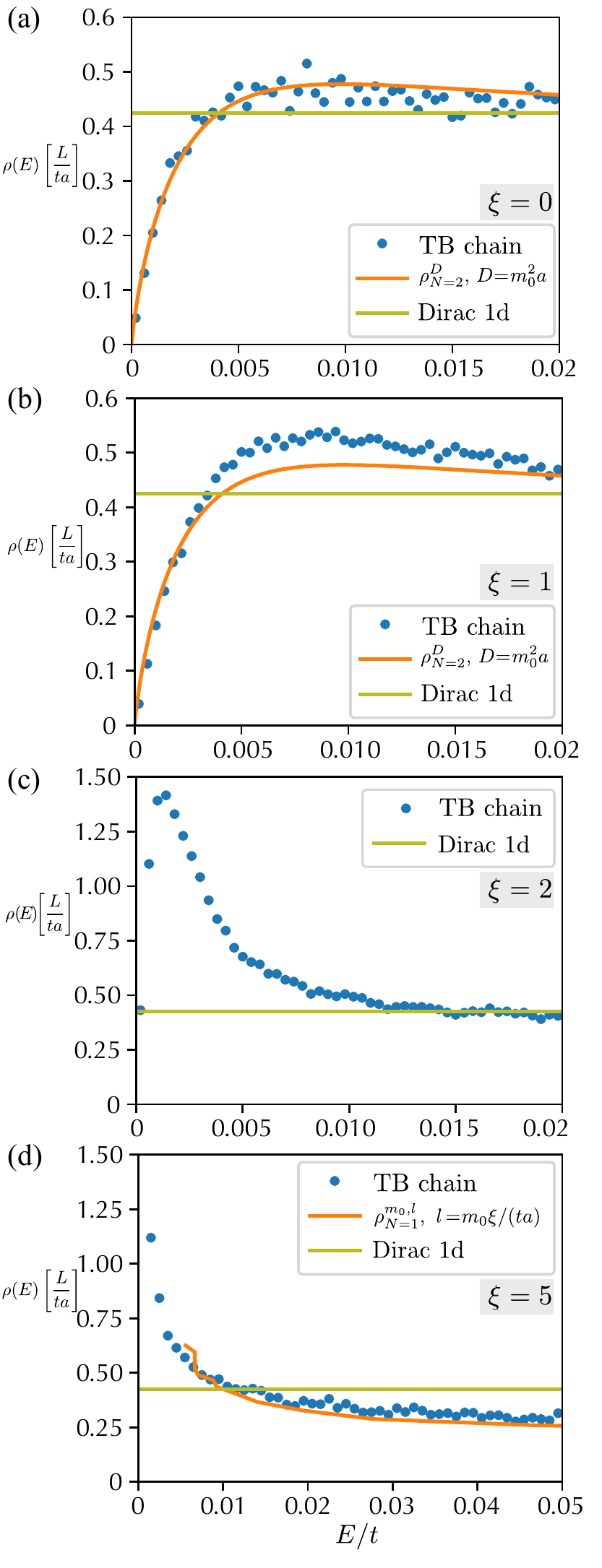}
\caption{\label{fig:masscut2} Crossover in DOS due to change with $\xi/a$ of the IVS in the two-valley model. We fix $m_0=0.2t$, and increase the correlation length (a slanted line-cut in Fig.~\ref{fig:mass}). (a) At $\xi/a=0$, $l_{latt}=0$, we find a surprising agreement with $\rho^{D=m_0^2 a}_{N=2}(E)$ (orange line, also in panel b). (b) At $\xi/a=1$, $l_{latt}=0.2$ and (c) at $\xi/a=2$, $l_{latt}=0.4$ show new crossover DOS shapes with the hump growing, narrowing, and moving towards $E=0$. (d) $\xi/a=5$, $l_{latt}=1$. The pseudogap is no longer discernible, and we recover the IVS-free DOS $\rho^{\mu=m_0,l=1}_{N=1}(E)$ (orange line).}
\end{figure}

Fig.~\ref{fig:masscut2} illustrates a cut through the phase diagram at fixed $m_0=0.2t$, hence on a slanted line through the crossover. In the crossover regime, which in the phase diagram forms a strip that is between $1<\xi/a<3$ and increases with $m_0\xi/(ta)$, the DOS curves develop extreme shapes, exemplified in Fig.~\ref{fig:masscut2}b,c, and don't match any known DOS curves. Importantly, as $\xi/a$ increases, the hump of the DOS curve increases rapidly in height, narrows, and moves towards lower energy, all the while the DOS vanishes at $E=0$. Finally, this hump becomes numerically indistinguishable from a sharp peak at $E=0$, representing the Dyson peak of $\rho^{\mu,l\approx0}_{N=1}(E)$, Fig.~\ref{fig:masscut2}d. This crossover in DOS indicates that either the value $\rho(E=0)$ jumps from zero to infinite at some value of $\xi/a$, or instead $\rho(E=0)=0$ at the center of the diverging Dyson peak.\footnote{Interestingly, a similar dilemma about how the Dyson peak arises as $l$ evolves from infinity to zero in $\rho^{\mu,l}_{N=1}(E)$ was discussed in Ref.\cite{BartoschComment}, where the hump plays no role, but the peak appears from the vanishing DOS close to $E=0$. The $\rho(E=0)$ in that case was related to boundary conditions for disorder. Note, by construction our disordered mass on the lattice obeys PBC in the $x$-direction.}
%%%%%%%%%%%%%%%%%%%%%%%%%%%%%%%%%%%%%%%%%%%%%%%%%%%
%%%%%%%%%%%%%%%%%%%%%%%%%%%%%%%%%%%%%%%%%%%%%%%%%%%

\begin{figure}\centering\includegraphics[width=0.45\textwidth]{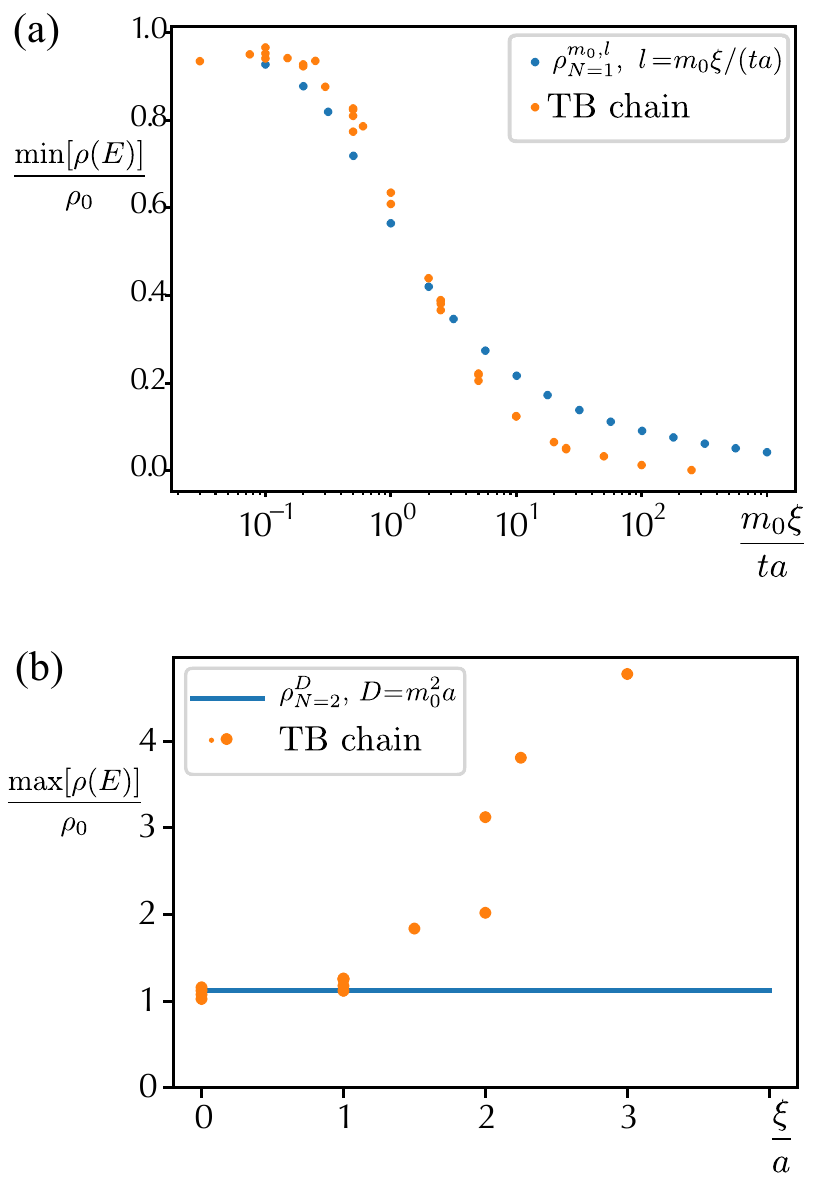}
\caption{\label{fig:compareDOS} Comparison of DOS of the two-valley system throughout the phase diagram to known DOS functions. The constant value of disorder-free low-energy DOS is $\rho_0$. (a) The minimum value of the two-valley DOS (orange dots) sorted by the value of $l$. Only DOS whose peak position is indistinguishable from $E=0$ are used. At each $l$ the multiple values correspond to multiple values of $\xi/a$, but this creates negligible variation, confirming the irrelevance of IVS. Comparison is to the dip of the known $N=1$ DOS (blue dots). (b) The maximum value of the two-valley DOS (orange dots) sorted by the value of $\xi/a$. Only DOS whose peak position is away from $E=0$ are used. At each $\xi/a$ the multiple values correspond to multiple values of $l$, and we note some variation, indicating the non-trivial interplay of IVS and the $l$ parameter. Comparison is to the (constant) value of the hump in the known $N=2$ DOS (blue line).}
\end{figure}

To better quantify the DOS of our two-valley system with respect to the known DOS functions, especially through the crossover, in Fig.~\ref{fig:compareDOS} we again split all numerical DOS into two classes depending on whether they have a peak at $E=0$ (class comparable to $N=1$) or at finite $E$ (class comparable to $N=2$), just like in the phase diagram Fig.~\ref{fig:mass}. In Fig.~\ref{fig:compareDOS}a we collect the DOS in class $N=1$ at any $\xi/a$ according to the value of their $l\equiv m_0\xi/(ta)$, and plot the depth of the dip (the minimum of DOS in units of the disorder-free DOS), overlaying it with the plot for the known DOS $\rho^{m_0,l}_{N=1}(E)$. We find a very good agreement for $l<4$, and a good agreement for large $l$. The plots suggest that introducing a weak dependence of the scaling factor $f_l$ on $l$ (see Eq.~\eqref{eq:param_map}), i.e., stretching the horizontal axis, could bring the minima of DOS into excellent agreement. In Fig.~\ref{fig:compareDOS}b we collect the DOS in class $N=2$ at any $l$ according to their $\xi/a$, and plot the height of the hump (the maximum of DOS  in units of the disorder-free DOS), overlaying it with the known value of hump of $\rho^{D=m_0^2 a}_{N=2}(E)$. We see the surprising match for white noise, which sharply worsens as $\xi/a$ grows and the IVS is reduced. The rising values for our DOS demonstrate the crossover in which the hump grows and moves towards zero energy. It is interesting to note that for much of the regime of strong IVS, i.e., most of the range $\xi/a\lesssim3$, the DOS has a strongly varying crossover shape, however once the peak moves to $E=0$, the DOS already excellently matches the $N=1$ result. This indicates that the crossover between $N=2$ and $N=1$ with increasing $\xi/a$ monotonously moves from high to low energies.
%%%%%%%%%%%%%%%%%%%%%%%%%%%%%%%%%%%%%%%%%%%%%%%%%%%
%%%%%%%%%%%%%%%%%%%%%%%%%%%%%%%%%%%%%%%%%%%%%%%%%%%

\section{Disordered next-nearest neighbor hopping and two valleys}
\label{sec:SOC}
In the previous section we emphasized the transition from two uncoupled valleys to a two coupled ones due to inter-valley scattering terms appearing when the correlation length of the disordered Semenoff mass reaches the lattice scale. The Semenoff mass term is on-site, which is in contrast to chiral quasi-1d systems for which the disordered term is usually a hopping term.
In our two-valley model based on spinless graphene there is however another gapping term which does represent (next-nearest neighbor) hopping, namely the Haldane mass. Such a term is generally expected to suppress IVS, but as we show the hopping can have a more complex intra-unit-cell structure so that the amount of IVS can be tuned. In this section we hence repeat the analysis of the phase diagram based on DOS for the two-valley system with disordered next-nearest-neighbor (NNN) hopping, but with a more detailed analysis of IVS.

\subsection{Lattice model}
\label{sec:dis_soc_model_latt}
Our tight-binding Hamiltonian with NNN hopping (and without Semenoff mass) is:
\begin{align}
  	\label{eq:lattice_Hamiltonian_SOC}
 H_\lambda^{TB} =& -t\displaystyle\sum_{<R\alpha,R'\alpha'>}c_{R\alpha}^\dagger c_{R'\alpha'}\\\notag
  &-\frac{i}{3\sqrt{3}}\sum_{R,\alpha,j}(-1)^\alpha\lambda_j(R)c_{R\alpha}^\dagger c_{R+a_j,\alpha} +H.c., \\\notag
\end{align}
where $R+a_j$, $j=1,2,3$, denotes the unit-cell translated by $\vec{a}_j$ w.r.t. to the unit-cell labeled by $R$, with PBC applied (see Fig.~\ref{fig:1}b). We assign the numerical values $\alpha=0,1$ to the sublattices $\alpha=A,B$. Importantly, due to the small width of our system, each lattice site $R\alpha$ is associated with only \emph{two} independent NNN bonds, say $\lambda_{1,2}(R)$, because the third associated bond, $\lambda_{3}(R)$, that would exist in infinite graphene is here identical to $\lambda_2(R)$ by PBC along the $y$-axis (see Fig.~\ref{fig:1}b). Nevertheless, we retain $\lambda_3$ as a free parameter to make calculations more transparent, while being aware that the terms $i=2$ and $i=3$ of the Hamiltonian Eq.~\eqref{eq:lattice_Hamiltonian_SOC} produce identical hopping terms between $R\alpha$ and $R-\vec{a}_1+\vec{a}_2,\alpha$ whose total amplitude is therefore $\lambda_2(R)+\lambda_3(R)\equiv\lambda_2'(R)$.

The lattice model Eq.~\eqref{eq:lattice_Hamiltonian_SOC} has the chiral symmetry $\Sigma:\,c_{RA}\rightarrow i c_{RB},\,c_{RB}\rightarrow -i c_{RA}$ (same as the Semenoff mass model) thanks to the quasi-1d periodicity in $y$-direction. This model with imaginary hoppings has a non-trivial spinless (pseudo)-time-reversal symmetry $T=U K$, with the unitary operation $U:\,c_{RA}\rightarrow  c_{RB},\,c_{RB}\rightarrow c_{RA}$, and is hence in the $BDI$ class. The $T$ symmetry also arises only because the quasi-1d nature allows $U$ to preserve NN hoppings and reverse the imaginary NNN hoppings. Note that our model in the homogeneous and locally (at each site) $C_3$-symmetric case $\lambda_1(R)\equiv \lambda_2(R)\equiv \lambda_3(R)\equiv \overline{\lambda_H}$ (hence, $\lambda_2'(R)\equiv 2\overline{\lambda_H}$) becomes the Haldane model\cite{Haldane} (thanks to the opposite sign of $\lambda$ on $A$ and $B$ sublattices) which in fully-2d graphene does break time-reversal symmetry.

A key assumption in our model of disorder is that the NNN hoppings $\lambda_{1,2,3}(R)$ have one fixed ratio within all unit-cells, defining a local intra-unit-cell configuration. From one unit-cell to another, their overall amplitude may vary arbitrarily:
\begin{equation}
  \label{eq:iuc}
\lambda_i(R)\equiv\lambda_i u(R),
  \end{equation}
with $u(R)$ a real function.

To interpret this intra-unit-cell structure of NNN hoppings that was absent in case of Semenoff mass, we refer to the homogeneous situation $u(R)\equiv1$, where the lowest order continuum theory,
  \begin{align}
    \label{eq:Hdir}
    \mathcal{H}_\lambda^{hom}&=\hbar v_F\tz\sx k_x+\overline{\lambda_{H}}\tz\sz\\\notag
    &-\frac{1}{3\sqrt{3}}\hat{e}_x\cdot\left(\sum_{j=1}^3\vec{a}_j \lambda_j\right)\sz k_x=\\\notag
&=\tz\sx k_x+\overline{\lambda_{H}}\tz\sz+\frac{a}{3\sqrt{3}}\left(\lambda_1-\frac{1}{2}\lambda'_2\right)\sz k_x
  \end{align}
with the third lattice vector $\vec{a}_3\equiv-\vec{a}_1-\vec{a}_2$, and with the homogeneous coupling
  \begin{equation}
    \label{eq:Hdirsoc}
\overline{\lambda_{H}}=\frac{1}{3}\sum_{j=1}^3\lambda_j.
  \end{equation}
We recover the continuum Haldane model with coupling $\overline{\lambda_{H}}$ only for the locally $C_3$-symmetric intra-unit-cell configuration $\lambda_1=\lambda_2=\lambda_3$, i.e., $\lambda_1=\frac{1}{2}\lambda_2'$, because this is equivalent to $\sum_j\vec{a}_j\lambda_j=0$ (note, the $\vec{a}_j$ sum to zero). To be more precise, the local (site-centered) $C_3$ symmetry means that in any given unit-cell, all 3 NNN hoppings emanating from site $A$ have the same value (and this is also true for the $B$ site). (Of course, only two of these are independent hoppings, namely $\lambda_1$, $\lambda'_2$ in that unit-cell.) Any breaking of this local symmetry induces corrections (very last term of Eq.~\eqref{eq:Hdir}).

The real-valued function $u(R)$ we assume to have a finite correlation length $\xi$ and a unit strength, while its average value is fixed to zero throughout this paper:
\begin{align}
  \label{eq:socdis}
  &\langle u(R)u(R')\rangle=\frac{a}{\sqrt{2\pi}\xi}\exp{\left[-\frac{(R-R')^2}{2\xi^2}\right]},\\
  &\langle u(R)\rangle=0.
\end{align}

We now proceed to take the low-energy continuum limit including the disorder.

\subsection{Low-energy one-dimensional Dirac model}
\label{sec:dis_soc_model_cont}
We derive the low energy continuum theory in real space by inverse Fourier transforming the small momenta around the Dirac points (Eq.~\eqref{eq:lattice_Hamiltonian_SOC}), which yields:
\begin{align}
  \label{eq:Hcontsoc}
  &\mathcal{H}_\lambda^{cont}= \hbar v_F\tz\sx(-i\partial_x)+\sum_{j=1}^3 \lambda_j\overline{u}(x)\tz\sz-\\\notag
&-\frac{\left(\vec{a}_j\lambda_j\right)_x}{3\sqrt{3}}\left[\frac{1}{2}\partial_x\overline{u}(x)+\overline{u}(x)i\partial_x\right]\sz-\\\notag &-\frac{\left(\vec{a}_j\lambda_j\right)_x}{3\sqrt{3}}\left[\frac{1}{2}\partial_x\hat{u}(x)+\hat{u}(x)i\partial_x\right]\tau_{x,y}\sz+\\\notag
&+O_2(a_{j,x}\partial_x)
%&(a_{j,x})^2\lambda_j\hat{u}(x)\partial_x^2 \tau_{x,y}\sz,
    \end{align}
where the notation is the same as in the mass case, Eq.~\eqref{eq:Hmlattice}, while the symbol $O_2(g)$ indicates quadratic- and higher-order terms in $g$. The first and second line of $\mathcal{H}_\lambda^{cont}$ describe the disorder-free energy and the intra-valley scattering due to a spatially varying NNN hopping on the lattice.

Importantly, and in contrast to the case of Semenoff mass, here the zero-th order IVS term ($\propto\hat{u}(x)\tau_{x,y}$) vanishes because the NNN hopping forbids backscattering. The forbidden backscattering is well-known for graphene with Kane-Mele spin-orbit coupling, which is a spinful model with two copies of the Haldane model, and we emphasize that our NNN hopping model is a spatially varying version of the Haldane model. Further, the first-order IVS term (third line of $\mathcal{H}_\lambda^{cont}$) vanishes for a special local site-centered-$C_3$-symmetric intra-unit-cell structure $\lambda_j\equiv\lambda$ (note, this is equivalent to $\sum_j\vec{a}_j\lambda_j=0$). In this case the first non-vanishing term is of third-order in $a_{j,x}\partial_x$. In Appendix~\ref{app:B} we detail the calculation of IVS in momentum space, and we show that for even more general disorder-types than Eq.~\eqref{eq:socdis} still the locally-$C_3$-symmetric configuration is the only one for which the first-order IVS term in the continuum vanishes. Hence, in case of NNN hopping disorder, the IVS is controlled by both the Fourier components of $u(R)$ (induced when $\xi$ is comparable to $a$ (Eq.~\eqref{eq:socdis})), but also by the intra-unit-cell ratio of the bonds $\lambda_1/\lambda'_2$, although the IVS is always at higher-order in continuum theory than for the disordered Semenoff mass.

The symmetry properties of the low energy model Eq.~\eqref{eq:Hcontsoc} are also different than for the case of Semenoff mass. There is the chiral symmetry $\Sigma=\sy$ due to the quasi-1d setup. Next, when the two valleys are decoupled, $\hat{u}(x)\equiv0$, then $\tau_z$ is a unitary symmetry, and each independent valley now depends on the intra-unit-cell structure of the NNN hopping: For generic $\lambda_1\neq\lambda_2$ there are no further symmetries, so each valley is in the chiral class $AIII$, while for the locally-$C_3$-symmetric disorder $\lambda_1\equiv\lambda_2$ there is effective time-reversal $T=\sigma_z K$ and the class is $BDI$. 
%the effective time-reversal symmetry $T=\sz K$, so each valley is in class $BDI$.
 In presence of IVS, $\hat{u}(x)\neq 0$, there are no unitary symmetries, and an effective spinless time-reversal symmetry of two valleys is inherited from the lattice: $T=\tau_x\sigma_xK$, which obeys $T^2=1$, and commutes with $\Sigma$, giving the chiral class $BDI$, as for the full lattice model. As $\xi/a$ grows, the one-dimensional Dirac equation is hence always in the chiral class $BDI$ $N=1,2$ for the locally-$C_3$-symmetric disorder, while it crosses over from $N=2$ $BDI$ to two copies of $N=1$ $AIII$ for generic intra-unit-cell disorders.

In Appendix~\ref{app:A} we apply transfer-matrix theory and find that the NNN model 
for a locally-$C_3$-symmetric disorder (for which IVS is  additionally suppressed compared to requirements of vanishing backscattering), indeed does not have the exponentially decaying conductance typical of chiral $N=2$ systems and the disordered Semenoff mass.
Instead, it behaves as an $N=1$ chiral model without IVS. Unfortunately this method becomes much more involved for generic NNN disorder $\lambda_1\neq\frac{1}{2}\lambda'_2$, and the calculation of conductance for the intermediate case of IVS in case of generic intra-unit-cell NNN hopping disorder is intricate. Therefore, we seek numerically the implications on DOS of the different degrees of IVS by studying any intra-unit-cell structure.

\section{Numerical DOS for disordered NNN-hoppings on lattice}
\label{sec:dis_soc_num}
%%%%%%%%%%%%%%%%%%%%%%%%%%%%%%%%%%%%%%%%%%%%%%%%%%%
\begin{figure}\centering\includegraphics[width=0.4\textwidth]{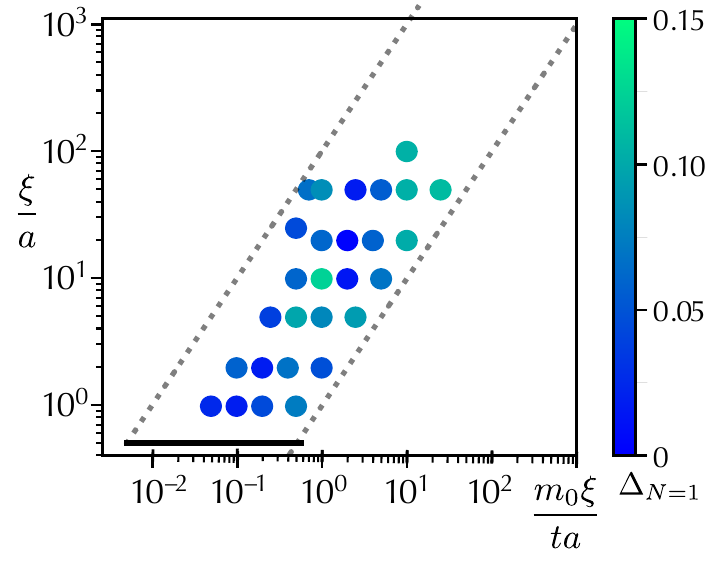}
\caption{\label{fig:soc}Phase diagram of the two-valley system with disordered NNN hoppings. The green-to-blue scale roughly measures the high-to-low agreement with $\rho^{\mu,l=0}_{N=1}(E)$ based on the depth of the dip in DOS. The DOS curves match well $\rho^{\mu,l}_{N=1}(E)$ throughout the entire phase diagram. The dotted lines show the limits of our numerical calculations, while the thick line marks that all values $\xi/a\leq0.5$ produce essentially the same lattice disorder with zero correlation length.}
\end{figure}
%%%%%%%%%%%%%%%%%%%%%%%%%%%%%%%%%%%%%%%%%%%%%%%%%%%

The analysis of the continuum theory implies that IVS is much weaker for disordered NNN hopping than for disordered Semenoff mass, however the IVS depends on the intra-unit-cell structure. The locally-$C_3$-symmetric disorder is given by $\lambda_1=\lambda_2=\lambda_3\Rightarrow\lambda_1=\frac{1}{2}\lambda_2'$, and has an additional suppression of IVS.

Somewhat surprisingly, with our numerical method we do not find significant difference in DOS curves obtained for various ratios $\lambda_1/\lambda_2'$. In Figure \ref{fig:soc} we show the representative phase diagram for $\lambda_1=\frac{1}{2}\lambda_2'\equiv\lambda_0$. From the high $\xi/a$ regime all the way down to lattice white noise, we do not find a sign of crossover, the DOS always matches well the decoupled-valleys DOS $\rho^{\mu,l}_{N=1}(E)$.

The question arises if the absence of IVS effects is real or an artifact of the numerical limitations. In Appendix~\ref{app:C} we roughly estimate the lengthscale $\xi_{IVS}$ on which inter-valley scattering becomes appreciable, as function of the intra-unit-cell deviation from the local-$C_3$-symmetric configuration, $\Delta\lambda/t\equiv(\lambda_1-\lambda_2'/2)/t$. We find that our system sizes $N_x\propto 10^4-10^5$ are larger than the estimate $\xi_{IVS}\propto10^3$ at least for relevant values $\Delta\lambda$ away from zero and at energies comparable to the dip feature. (However, the estimated IVS lengthscale at those energies becomes comparable to our largest $N_x$ for $\Delta\lambda=0$.) Hence even when our finite tight-binding models should be able to capture IVS effects, these do not seem to be present. In Appendix~\ref{app:C} we substantiate this analysis through a perturbative calculation of the self-energy in the continuum, which confirms that for any system size and value of $\Delta\lambda/t$ there is always a window of small energies in which the effect of IVS is negligible, hence the Dyson peak is expected to persist at any $\xi/a$.

\section{Discussion and Conclusions}
We have shown that in chiral quasi-1d systems with two valleys (i.e., two crossings at different finite momenta at low-energy) the disorders in each valley become intrinsically inter-dependent. We study the crossover in between decoupled and coupled valleys as the correlation length of disorder approaches the lattice constant, showing a non-trivial evolution of DOS between a divergent peak and a pseudogap, while the system becomes insulating. We find that the inter-dependent disorder has at least quantitative effects on the DOS in the regime of coupled valleys even in the limit of zero correlation length for disorder. In the case of disorder in NNN hoppings, we find that the amount of IVS can be tuned to appear at different orders in the low-energy theory, but the lowest-energy DOS always behaves as if the valleys are decoupled throughout the phase diagram.

The first question one could ask is how universal our results are. It is known that in chiral quasi-1d systems there are non-universal corrections and an additional parameter may be needed to describe the disorder\cite{BMF00NPB,BMF01physE}, although this does not include the specific properties of our disorder. Clearly, two-valley systems are generic, but what might be less so is the fact that we had a chiral system with strong IVS, as happened in the case of disorder in Semenoff mass which is an on-site term. An interesting test-case could be the versions of Haldane model with real NNN hoppings,\cite{Haldane} instead of our case of purely imaginary NNN hoppings, as the backscattering may be allowed and IVS may not be strongly suppressed.

Staying with the current graphene-based systems, physical generic disorder breaks all lattice symmetries, so it is somewhat artificial that we considered NNN hoppings with strictly opposite sign on the two sublattices. Our motivation was to focus on the gap-opening Haldane coupling in the continuum limit, but there are other gap-opening possibilities with hopping disorder, for example, simultaneous presence of valley-Zeeman and Rashba spin-orbit coupling (in the spinfull system). We note that valley-Zeeman spin-orbit NNN hopping in its spinless version is complementary to the Haldane mass, together they allow to disentangle the NNN hoppings on two sublattices. However the valley-Zeeman-type NNN hopping is not gap-opening, its disordered version is given by identity operator in sublattice space and hence it is extremely ineffective in influencing the DOS.

It would be however interesting to study multi-valley models not based on graphene, because the chirality we have is fully dependent on our quasi-1d setup, i.e., setting $k_y=0$ to have $\sy$ as the chiral operator. This limits us to only $N=2$ in the lattice model. In principle, a good strategy could be to construct multi-valley Dirac models with chirality and back-engineer the tight-binding lattice model. Generalizations of the fluctuating gap model may also be envisaged\cite{BartoschThesis}.

Technically speaking, our work poses challenges that are quite general. One is the extension of analytical techniques for calculating DOS\cite{GrabschThesis,BartoschAN} to the case of inter-dependent disorder, even in the simplest case of continuum white noise, where the non-trivial inter-dependence is that the disorder matrix is not isotropic in $N\times N$ space. Interestingly, the standard transfer-matrix approach to calculating conductance also seems to become analytically intractable for the case of generic NNN hopping disorder in our complex four-component representation of the Dirac equation.

We also emphasize that the theoretical treatment of our problem is intrinsically challenging as it exactly plays on lattice effects in the low-energy limit, which are not simply questions of cutoff scales, singularity regularizations or boundary conditions. In particular, the lattice model when $\xi$ becomes comparable to $a$ and the two valleys become coupled is not a problem with weak IVS, and it is not straightforwardly taken to the continuum limit neither by $\xi/a\rightarrow0$ (and then connecting lattice and continuum white noises) nor by $\xi/a\rightarrow\infty$ (and then smoothing the disorder function). Hence our low-energy Dirac theory analysis should be questionable in the crossover regime, and obviously the non-trivial interpolating DOS function found numerically should motivate new approaches to these types of systems.

\acknowledgments
We thank A. Grabsch for instructing us on the technique of deriving the $N=2$ DOS in Ref.\cite{Grabsch}, and for discussing the possibilities for generalizing the method. We also thank J. Bardarson and G. Montambaux for useful discussions.

\appendix

\section{Conductance of quasi-1d models of disordered mass and NNN hoppings}
\label{app:A}
Based on the continuum models $H^{cont}_m$ and $H^{cont}_{\lambda}$ in Eqs.~\eqref{eq:Hmcont},~\eqref{eq:Hcontsoc}, we consider simplified Hamiltonians of the same structure and derive the conductance. In case of NNN hopping, we are only able to treat the locally-$C_3$-symmetric configuration $\lambda_1=\frac{1}{2}\lambda_2'$.

We begin with the Semenoff mass:
\begin{equation}
\mathcal{H}=iv_F\partial_x\sigma_x\tau_z + m_1(x)\sigma_z + m_2(x)\tau_x\sigma_z + m_3(x)\tau_y\sigma_z
\end{equation}
where $m_1$ and $\sqrt{m_2^2+m_3^2}$ are respectively the low and high momentum components of the original mass field.

This Hamiltonian has $\sigma_y$ as a chiral symmetry. So, for the sake of simplicity, we rotate into the eigenbasis of $\sigma_y$ (so $\sigma_y \to \sigma_z \to -\sigma_y$ and $\sigma_x$ is invariant). Consequently, the scattering matrix is
\begin{align}
S=T_x \exp\Biggl(\frac{1}{v_F}\int_0^{\delta_L} &m_1(x)\tau_z\sigma_z+im_2(x)\tau_y\sigma_z \notag\\
											   &-im_3(x)\tau_x\sigma_z \, \mathrm{d}x \Biggr)
\end{align}

Hence we have the following symmetries for $S$:
\begin{align*}
\sigma_zS\sigma_z=S \\
\tau_z\sigma_yS\tau_z\sigma_yS^\dagger(=\tau_z\sigma_xS\tau_z\sigma_xS^\dagger)=
\mathbb{1}
\end{align*}
The expression in parentheses is simply deduced from the chiral symmetry and the first current, so it doesn't bring any additional information.

As a consequence, in the basis $\{\ket{1,1},\ket{1,-1},\ket{-1,1},\ket{-1,-1}\}$, where the first number is the $\sigma_z$ eigenvalue and the second one the $\tau_z$ eigenvalue, we can parametrize the $S$ matrix in the following way:
\[S=\begin{pmatrix}
M & \\
&\tau_zM^{\dagger -1}\tau_z
\end{pmatrix},\]
where $M$ is a $2\times 2$ complex invertible matrix.

We have checked the two moments of the increment of the $x_j$ variable as introduced in Ref.\cite{Brouwer98N} As we have $N=2$, we get the following results:
\begin{align*}
<\delta x_i> &= \frac{2\sigma^2\delta L}{3v_F^2}\coth(x_i-x_j) \\
<\delta x_i^2> &= \frac{\sigma^2\delta L}{3v_F^2} \\
<\delta x_i \delta x_j> &= -\frac{\sigma^2\delta L}{2v_F^2}
\end{align*}
where $\sigma$ is the standard deviation of the mass field and $j\neq i$.

Given that the cross correlation between the increment of the transmission eigenvalues is negative and that the interaction between the two of them is negative, there is no competition in the diffusive process and the eigenvalues should crystallize at extensive opposite values. Thus the conductivity exponentially decays.

Now we move to the NNN hopping model, Eq.~\eqref{eq:Hcontsoc}. We only modify the intra-valley term from $m\sz$ to $m\sz\tz$, and hence the simplified model is relevant only for the special locally-$C_3$-symmetric case $\sum_j\vec{a}_j\lambda_j=0$. The scattering matrix is now written
\begin{align}
S=T_x \exp\Biggl(\frac{1}{v_F}\int_0^{\delta_L} &m_1(x)\sigma_z+im_2(x)\tau_y\sigma_z \notag\\
											   &-im_3(x)\tau_x\sigma_z \, \mathrm{d}x \Biggr) \, .
\end{align}

We have the same chiral symmetry and conserved currents and consequently the same parametrisation of the S matrix. However the average displacement of the eigenvalues of S induced by an infinitesimal slice of width $\delta L$ is now 0. There is no drift of the eigenvalues of $M$. Consequently, instead of a conductance which decays exponentially with the sample size, we have fluctuations\cite{Brouwer98N}.

\section{Inter-valley scattering due to intra-unit-cell structure of NNN hopping}
\label{app:B}
We now examine the effect of different choices for the intra-unit-cell structure $\lambda_{1,2,3}$ of NNN hopping, finding that they greatly influence the scattering for a given spatial variation $u(R)$ between unit-cells labeled by $R$. We emphasize again that in our quasi-1d system there are in fact only two independent NNN couplings, $\lambda_{1}$ and $\lambda_@'\equiv\lambda_{2}+\lambda_{3}$). In particular the central question in this Appendix is: How does the intra-unit-cell structure affect the inter-valley scattering (IVS), given that the inter-unit-cell variation $u(R)$ may have large scattering momenta that connect the valleys?

For easier reading we will treat momenta as two-dimensional, and implement the quasi-1d nature of the system ($k_y\equiv0$) at the end of calculation. To understand the scattering, we start from the momentum space version of Eq.~\eqref{eq:lattice_Hamiltonian_SOC}:
\begin{widetext}
\begin{equation}\label{eq:Hlattk}
H = -t\displaystyle\sum_{\vec{k}} c_{\vec{k}A}^\dagger\gamma(\vec{k})c_{\vec{k}B}+H.c.
+
\sum_{\vec{k}_1,\vec{k}_2} c_{\vec{k}_2\alpha}^\dagger\left[
\frac{1}{3\sqrt{3}}\tilde{u}(\vec{k}_2-\vec{k}_1)\Gamma_\lambda(\vec{k}_1,\vec{k}_2)\sz^{\alpha\beta}\right] c_{\vec{k}_1\beta},
\end{equation}
\end{widetext}
where $\tilde{u}$ is the Fourier transform of the space-dependent term $u(R)$, while the intra-unit-cell factors are
\begin{align}\notag
&\gamma(\vec{k})=\sum_{j=1}^3\exp(-i\vec{k}\cdot\vec{d}_j)\\
&
\begin{aligned}
  \Gamma_\lambda(\vec{k}_1,\vec{k}_2)
  &= i\sum_{j=1}^3 \lambda_j\left[\exp(i\vec{k}_1.\vec{a}_j)-\exp(-i\vec{k}_2.\vec{a}_j)\right]\\
  &= -2\sum_{j=1}^3 \lambda_j\sin(\vec{k}_+.\vec{a}_j)\exp(i\vec{k}_-\cdot\vec{a}_j),
\end{aligned}
\end{align}
with the three NN bond vectors $\vec{d}_j$, and the auxiliary momenta $\vec{k}_\pm\equiv(\vec{k}_1\pm\vec{k}_2)/2$. The fact that the NNN part of the Hamiltonian is purely imaginary causes $\Gamma_\lambda(\vec{k}_1,\vec{k}_2)$ to be odd in $\vec{k}_+$, with the consequence that exact backscattering $\vec{k}_2=-\vec{k}_1$ is forbidden, i.e., $\Gamma_\lambda(\vec{k}_+=0)=0$. This makes the spatially varying NNN hopping \emph{a priori} less efficient in scattering between valleys at $\pm K$ than the varying mass, the latter having a trivial intra-unit-cell factor. The effect of the intra-unit-cell structure of NNN hopping is revealed in the low energy theory: we label the Dirac momenta $\pm\vec{K} =\pm \frac{4\pi}{3\sqrt{3}a}\hat{e}_x$, with $K\equiv|\vec{K}|$, as the valleys $\tz=\pm$, and consider momenta in their vicinity, in particular to study IVS we take two unrelated momenta in opposite valleys $\vec{k}_1=\taus \vec{K}-\vec{q}_0$, $\vec{k}_2=-\taus\vec{K}+\vec{q}_0+\vec{q}$, and we get the value of intra-unit-cell factor
\begin{align}\notag\label{eq:Gamma}
\Gamma^{\taus}_\lambda(\vec{k}_1,\vec{k}_2)
  &= i z^\taus\sum_{j=1}^3 \lambda_j\left[\exp(-i\vec{q}_0.\vec{a}_j)-\exp(-i(\vec{q}_0+\vec{q})\cdot\vec{a}_j)\right]\\
  &\approx -z^\taus\sum_{j=1}^3 \left(\vec{a}_j\lambda_j\right)\cdot\vec{q}+O(q^2,q_0^2,q_0 q),
\end{align}
where $z\equiv\exp(-i2\pi/3)$. We now see that the lowest order term in IVS can be exactly cancelled if the intra-unit-cell structure obeys $\sum_{j=1}^3 \vec{a}_j\lambda_j=\vec{0}$, which in our quasi-1d system gives $\lambda_2+\lambda_3$ upon projection on $y$-axis (a trivial condition since in quasi-1d only $\lambda_2'=\lambda_2+\lambda_3$ exists), and $\lambda_1-\frac{1}{2}(\lambda_2+\lambda_3)=0$, i.e., $\lambda_1=\frac{1}{2}(\lambda_2'$ upon projection on $x$-axis. Hence, apart from the suppression of IVS due to lack of backscattering, there is an additional suppression of IVS for the locally-$C_3$-symmetric intra-unit-cell structure $\lambda_1\equiv \lambda_2\equiv \lambda_3\Rightarrow\lambda_1\equiv\frac{1}{2}\lambda_2'$.

To understand the generality of the suppression of IVS, we consider an even more general set of intra-unit-cell structures than based on the three NNN hoppings in Fig.~\ref{fig:1}b. Our physical motivation comes from the case of a dense random distribution of adatoms, when there is no natural partitioning of bonds into unit-cells. To illustrate this non-trivial ambiguity, first note that so far we assigned to a given site (say an $A$ site in unit-cell $R$), the three NNN $A-A$ hoppings emanating from this site. Imagine now the alternative of assigning to this site the three NNN $A-A$ bonds which form a triangle such that our site is at its center. When we set $\lambda_1(R)\equiv \lambda_2(R)\equiv \lambda_3(R)$ both choices produce a locally (at our site) site-centered-$C_3$-symmetric choice of $A-A$ bonds. Both choices might be natural for an adatom positioned directly on top of our site. However, in fact the second choice of intra-unit-cell structure does \textit{not} cancel the first order term in the IVS in $\Gamma_{\lambda}$ as the first choice does. Hence, to clarify the necessary conditions for the additional IVS suppression, we consider even more general structures: to each site, say our site $A$, we assign three arbitrary NNN $A-A$ bonds chosen from the set of bonds which either emanate from our site or are inside a hexagon plaquette which contains our site. Enumerating all possibilities and studying their $\Gamma_{\lambda}$, we find that \textit{only} our original choice of bonds emanating from the site as in Fig.~\ref{fig:1}b, with locally site-centered-$C_3$-symmetric values of NNN hoppings when $\lambda_1\equiv \lambda_2\equiv \lambda_3$, gives an exact canceling of first order term in $\Gamma_\lambda$ as in Eq.~\eqref{eq:Gamma}. Hence throughout the text for simplicity we stay with the intra-unit-cell structure based on assigning three bonds NNN hoppings $\lambda_{1,2,3}$ to a site just as in from Fig.~\ref{fig:1}b. then we are able to address the additionally suppressed IVS when $\lambda_1\equiv \lambda_2\equiv \lambda_3$.

\section{Estimate of inter-valley scattering lengthscale and its effects}
\label{app:C}
To quantify and interpret the suppression of IVS due to variation of intra-unit-cell configuration of disordered NNN hoppings, we proceed to extract a typical IVS length. This IVS lengthscale is useful to check that the numerical tight-binding results in Section~\ref{sec:SOC} could, in principle, demonstrate IVS effects since the system length $N_x$ is beyond the IVS lengthscale, at least for some intra-unit-cell configurations. To estimate the lengthscale we use a periodic inter-unit-cell spatial profile $u(R)$ with Fourier components at $2K$ being able to scatter between valleys, and this is done in the first subsection.

We then proceed to a continuum-white-noise profile of disorder and use a disorder-averaged self-energy to argue why there are no IVS effects at low energies in case of disordered NNN hoppings. We contrast the results to the case of disordered Semenoff mass.

\subsection{Inter-valley scattering lengthscale due to a periodic coupling}
\label{sec:fgr}
We first define a characteristic IVS time $\tau$ as the time it takes for an electron starting in state in one valley to end up in the other valley. Using the Fermi golden rule:
\begin{equation}
  \label{eq:fgr} \frac{1}{\tau_{\vec{k}}}=\sum_{|\vec{k'}|=|\vec{k}|}\Gamma_\lambda(\vec{K}+\vec{k'},-\vec{K}+\vec{k}),
\end{equation}
where energy conservation is apparent. Then we define a characteristic IVS length simply by $\xi_{\vec{k}} = v_F \tau_{\vec{k}}$.

We start from the special locally-$C_3$-symmetric intra-unit-cell NNN hopping configuration $\lambda_1\equiv \lambda_2\equiv \lambda_3\equiv\lambda_0$, and the resulting low energy continuum theory of Eq.~\eqref{eq:lattice_Hamiltonian_SOC}:
\begin{equation}
  \label{eq:fgrH}
\mathcal{H}_0=\hbar v_Fk_x\tz\sx+\lambda_0 u(x,T)\tz\sz\spz+\msem\, u(x,T)\sz,
\end{equation}
where the inter-unit-cell spatial dependence of couplings,
\begin{equation}
  \label{eq:fgrU}
  u(x,T)=\theta(T)e^{-\frac{x^2}{2d^2}}\cos(Qx),
\end{equation}
is: (1) chosen periodic in $x$ with wavevector $Q$ that scatters between valleys (while momentum $\vec{k}\cdot\vec{a}_2=0$ in our quasi-1d system), (2) smoothly confined on a lengthscale $d$ which is taken to equal the system length $d=L$ at the end of calculation\cite{BartoschComment}, and (3) switched on at time $T=0$ as usual in applying the Fermi golden rule.

Starting with NNN hopping by setting $\msem=0$, we find that states at one valley, say $k-=-K+q$ can only be scattered to the other valley if $Q=2K$ or $Q=2K-2q$, but the latter represents exact backscattering which is nullified by $\Gamma$ factor. Then, under the assumption that $|q|\ll K$, the rate of transition from $-K+q$ to $K+q$ is:
\begin{align}
  \label{eq:fgrQ}\notag
  \frac{1}{\tau_{Q}(q)} &= \frac{\lambda_0^2 d^2}{L^2}\frac{L}{v_F}\sin(qa)^6\cos(qa)^2 \approx \frac{\lambda_0^2d^2}{L^2}\frac{L}{v_F}(qa)^6\\
  &= \frac{\lambda_0^2 N}{t}\frac{d^2}{L^2}(qa)^6,
\end{align}
where in the last line we convert the continuum parameters to lattice parameters of graphene, and introduce the number of unit-cells in the system $L=N_x a$ (we drop inessential prefactors such as $\hbar$ and constants of order 1). We now take the limit $d\rightarrow L$ and get for the IVS length:
\begin{equation}
  \label{eq:fgrX}
  \xi_{Q,\lambda_0}(q) = \frac{v_F^2}{\lambda_0^2L}\frac{1}{(qa)^6}=\left(\frac{t}{\lambda_0}\right)^2 \frac{a}{N_x}\frac{1}{(qa)^6}.
\end{equation}

Now we can estimate the system size up to which the IVS is effectively absent:
\begin{equation}
  \label{eq:fgrnoIVS}
\xi_{Q,\lambda_0}(q) \gg L \Rightarrow N_x\ll \frac{t}{\lambda_0}\frac{1}{(qa)^3}.
\end{equation}
For a numerical estimate relevant to our tight-binding simulations, we take parameters $\lambda_0 = 0.1t$ and $qa= 0.1$, where the latter corresponds to the span of small momenta where Dirac approximation is valid. This gives a maximal system length
\begin{equation}
  \label{eq:fgrNmax}
N_{\mathrm{no-IVS}}^{Q,\lambda_0}=\frac{t}{\lambda_0}\frac{1}{(qa)^3}\approx 10^4.
\end{equation}
We remark that this estimate is consistent with the validity range of the Fermi golden rule calculation, which assumes weak scattering, so it is consistent to consider a IVS length larger than system size. %More formally, the scattering time sets an upper bound for the perturbative expansion. So this new upper bound also needs to be greater than the lower bound which is $1/\sigma$. So we need $\tau\sigma \gg 1$. But $\tau\propto 1/(\rho W^2)$ and in the limit $d\to L$, $\rho=1/\sigma$. So $\tau\sigma\gg 1$ is equivalent to $\sigma/W \gg 1$ which is imposed at the beginning. The fact that $\rho=1/\sigma$ is a consequence of the linear dispersion relation of clean graphene.
Finally, note that in the simulations we study a small low-energy range corresponding to $qa=0.01$ in the continuum Dirac dispersion, which formally increases the estimate of $N_{\mathrm{no-IVS}}$. Hence, our systems of length $N_x\propto 10^5$ should be able to exhibit IVS effects.

Now we consider deforming the intra-unit-cell structure of NNN hopping away from the special locally-$C_3$-symmetric configuration, by introducing a non-zero quantity
\begin{equation}
  \label{eq:fgrDL}
  \Delta\lambda = (\lambda_1-\lambda_2) + (\lambda_1-\lambda_3)=2\lambda_1-\lambda_2',
\end{equation}
and a continuum model as in Eq.~\eqref{eq:fgrH} with terms from Eq.~\eqref{eq:Hcontsoc}. The same procedure as above gives the contribution to scattering rate 
\begin{equation}
  \label{eq:fgrXDL}
  \frac{1}{v_F\tau_{Q,\Delta\lambda}(q)} = \left(\frac{\Delta\lambda}{t}\right)^2\frac{N_x}{a}(qa)^2,
\end{equation}
which means that at low energies the total IVS is strongly dominated by a finite $\Delta\lambda$ contribution (scaling as $q^2$) compared to the contribution of locally-$C_3$-symmetric configuration in Eq.~\eqref{eq:fgrX} (scaling as $q^6$). The maximal no-IVS system length in case of dominant contribution from $\Delta\lambda$ is
\begin{equation}
  \label{eq:fgrDNmax}
N_{\mathrm{no-IVS}}^{Q,\Delta\lambda}=\frac{t}{\Delta\lambda}\frac{1}{qa}\approx 10^3,
\end{equation}
with the numerical example of a small $\Delta\lambda=0.01t$ and $qa= 0.1$ still dominating the IVS compared to Eq.~\eqref{eq:fgrNmax}.

To put the results into context, we contrast them to the exact same calculation for the case of IVS due to a varying Semenoff mass parameter, i.e., $\lambda_0=0$ and $m\neq0$ in Eq.~\eqref{eq:fgrH}. We obtain the expected momentum-independent IVS:
\begin{align}
  \label{eq:fgrm}
&\frac{1}{\tau_\msem} = \frac{\msem^2L}{v_F} = \frac{\msem^2 N_x}{t}\\\notag
&\xi_{Q,\msem} = \frac{v_F^2}{\msem^2L} = \left(\frac{t}{\msem}\right)^2\frac{1}{N_x}a\\\notag
& N_{\mathrm{no-IVS}}^{Q,\msem}=\frac{t}{\msem}\approx10,
\end{align}
with the estimate of $N_{\mathrm{no-IVS}}$ for parameters $\msem=0.1t$ and $qa=0.1$ showing a great dominance of IVS due to disordered mass compared to due to any type of disordered NNN, as expected due to forbidden backscattering in the latter system.

It is important to clarify that our finding of suppression of IVS in the case of the locally-$C_3$-symmetric intra-unit-cell configuration of NNN hopping (compare Eqs.~\eqref{eq:fgrX} and \eqref{eq:fgrXDL}) is not a specialty of our quasi-one-dimensional system, but appears in the two-dimensional graphene sheet too.

\subsection{Self-energy analysis for continuum white noise couplings}
\label{sec:disav}
Having a random spatial variation from one unit-cell to the next should cause much more comprehensive scattering than the simple sinusoidal variation of momentum $Q=2K$ we discussed in Appendix~\ref{sec:fgr}. To quantify the effect of randomness in NNN hoppings on the IVS, we therefore consider the low energy model as in Eq.~\eqref{eq:fgrH} with continuum Gaussian white noise:
\begin{align}
  \label{eq:randcorr}
  &\left<u(x)u(x')\right>=a\,\delta(x-x'),\\\notag
  &\left<\lambda_i(x)\lambda_j(x')\right>=\lambda_0^2a\,\delta(x-x'),
\end{align}
where we focus only on the locally-$C_3$-symmetric intra-unit-cell configuration of NNN hoppings, $\lambda_1=\lambda_2=\lambda_3\equiv\lambda_0$. We calculate in the usual way the disorder-averaged Green's function to lowest order. The central quantity is the self-energy, given to lowest order by:
\begin{equation}
%\feynmandiagram[baseline=(i.base),large,horizontal=i to o]{
%i -- [edge label'={$\vec{q}$}] o,
%i -- [scalar, edge label = {$\vec{k}-\vec{q}$}] [star, star points = 6, star point ratio = 0.5, fill=black, inner %ep=5pt] --[scalar, edge label = {$\vec{q}-\vec{k}$}] o,
%};
\Sigma = \sum_{\vec{q}} \left<V(\vec{q},\vec{k})^\dagger G_0(\vec{q}) V(\vec{q},\vec{k})\right>,
\end{equation}
with $G_0$ the unperturbed (retarded) Green's function and $V$ representing the perturbation, i.e., our random NNN hopping term:
\[V(\vec{q},\vec{k}) = \hat{\lambda}(\vec{q},\vec{k}) = i\tilde{\lambda}(\vec{q}-\vec{k})\Gamma(\vec{q},\vec{k})\sz, \]
where $\Gamma$ is the geometric term as in Eq.~\eqref{eq:Gamma} but in the continuum limit.

The true self energy contains all scattering contributions, but for our purpose we will split it in two parts $\Sigma=\Sigma_i+\Sigma_o$, the intra-valley and inter-valley, respectively, and focus on IVS part $\Sigma_o$. When computing $\Sigma_o$, we choose momenta in opposite valleys, and simply redefine the $q$ and $k$ symbols: $q\rightarrow\mathbf{K}+q$ and $k\rightarrow-\mathbf{K}+k$, which are now small momenta, $q,k\ll K$. We note that the $\sigma_z$ structure of our NNN hopping will change the direction of movement when scattering inside a valley and preserve it when scattering to the other valley, and hence $\Sigma_o$ preserves $\tz\sx$ whose eigenvalue we label by $\sigma$. The following calculation is for right movers, but easily extended to left movers. Using that $q,k\ll K$, we get
\begin{align}
  \label{eq:disS}
  &|\Gamma(q,k)| = \frac{9}{4}|q^2-k^2|a^2+o(q^2+k^2),\\
  &\Sigma_o^{\sigma}=\sum_q \frac{\lambda_0^2}{N_x}\frac{81}{16}(q^2-k^2)^2a^4\frac{1}{\omega-v_F\sigma q+i0^+}.
\end{align}
Letting the microscopic lengthscale $a$ go to zero, we obtain the real and imaginary parts of the IVS self-energy:
\begin{align}
  \label{eq:3}
  &\Re(\Sigma_o)=-3\sqrt{3}\left(\frac{\lambda_0a}{v_F}\right)^2\omega,\\
  &\Im(\Sigma_o)= -\frac{81\sqrt{3}}{64}\frac{\lambda_0^2a}{v_F}\left(\frac{\omega^2}{v_F^2}-k^2\right)^2a^4,
\end{align}
which after resumming give the spectral function
\begin{align}
  \mathcal{A}(k,\omega) &= -\frac{1}{\pi}\Im(G(k,\omega))\\
  &= \frac{1}{\pi}\frac{Z^2\alpha(\omega^2-v_F^2k^2)}{(\omega-Zv_F\sigma k)^2+Z^2\alpha^2(\omega^2-v_F^2k^2)^4},
\end{align}
where we defined the disorder-dependent factors $Z^{-1} =1+3\sqrt{3}\left(\frac{\lambda_0a}{v_F}\right)^2$ and $\alpha = \frac{81\sqrt{3}}{64}\frac{\lambda_0^2a^5}{v_F^5}$.

The main result is that for a given strength of disorder $\lambda_0$ there is a range of small momenta around the Dirac point where the IVS keeps quasiparticles sharp. This result is in accord with the case of IVS due to a periodic spatial variation with momentum $2K$, Eq.~\eqref{eq:fgrX}, which showed that for any system size and amplitude of periodic NNN hopping there is a range of momenta for which the IVS lengthscale exceeds the system size.

To put the disorder averaging into context, we contrast it with the case of continuum white noise in Semenoff mass, for which the perturbation term in Eq.~\eqref{eq:fgrH} is given by:
\begin{align}
  \label{eq:2randm}
  &V(\vec{q},\vec{k})=\tilde{\msem}(\vec{q}-\vec{k})\sz,\\
  &\left<\msem(x)\msem(x')\right>=m_0^2a\,\delta(x-x').
\end{align}
The $\sigma_z$ structure of the mass will have the same effect as previously, i.e., preserving $\tau_z\sigma_x$ for the IVS part, and changing the sign for the intra-valley part. Same procedure as above yields:
\[\Sigma_{i/o} = -i\frac{am_0^2}{2v_F},\]
showing that both intra- and inter-valley scattering contribute equally to the broadening of states. Because there is no dependence on momenta and energy, we can straightforwardly extract an IVS lengthscale:
\begin{align}
  \label{eq:disXm}
&\tau_{\textrm{dis},m} = \frac{v_F}{am_0^2}=\frac{t}{m_0^2},\\
&\xi_{\textrm{dis},m} = \frac{v_F^2}{a m_0^2}=\left(\frac{t}{m_0}\right)^2a,
\end{align}
where we dropped inessential factors, and obtained in accordance with Eq.~\eqref{eq:fgrm} that for $m_0\lesssim t$ the IVS is significant even for very small systems.

\bibliography{jb_bib}
\end{document}